\documentclass[12pt,preprint]{aastex}
\usepackage{aas_macros}
\usepackage{graphicx}

\newcommand{\gpcmc}{\mathrm{g/cm^3}}
\newcommand{\udel}[1]{\partial_{#1}}
\newcommand{\tdel}[1]{\partial^{#1}}
\newcommand{\pdel}[2]%
{\frac{\partial{#2}}{\partial{#1}}}
\newcommand{\pddel}[2]%
{\frac{\partial^2{#2}}{\partial{#1}^2}}
\newcommand{\gauss}{\mathrm{G}}

\newcommand{\text}[1]{\rm #1}

\newcommand{\km}{\mathrm{km}}

\begin{document}

\title{Gravitational Wave Signatures of Magnetohydrodynamically-Driven
Core-Collapse Supernova Explosions}

\author{Tomoya Takiwaki\altaffilmark{1} and Kei Kotake\altaffilmark{1,2}}
\affil{\altaffilmark{1}Center for Computational Astrophysics, National Astronomical Observatory of Japan, 2-21-1, Osawa, Mitaka, Tokyo, 181-8588, Japan}
\affil{\altaffilmark{2}Division of Theoretical Astronomy, National Astronomical Observatory of Japan, 2-21-1, Osawa, Mitaka, Tokyo, 181-8588, Japan}
\email{takiwaki.tomoya@nao.ac.jp,kkotake@th.nao.ac.jp}
\begin{abstract}

By performing a series of two-dimensional, special relativistic magnetohydrodynamic (MHD)
 simulations, we study signatures of gravitational waves (GWs)
in the magnetohydrodynamically-driven core-collapse supernovae.
In order to extract the gravitational waveforms, we present
 a stress formula including contributions both from magnetic fields
 and special relativistic corrections.

By changing the precollapse magnetic fields and initial angular 
 momentum distributions parametrically, we compute twelve models. 
 As for the microphysics, 
a realistic equation of state is employed and the neutrino cooling is 
taken into account via a multiflavor neutrino leakage scheme.
 With these computations, we find that the total GW amplitudes 
show a monotonic increase after bounce for models with a strong precollapse 
magnetic field ($10^{12}$G) also with a rapid rotation imposed.
 We show that this trend stems both 
from the kinetic contribution of MHD outflows with large radial velocities and
 also from the magnetic contribution dominated by the toroidal magnetic fields 
 that predominantly trigger MHD explosions. 
For models with weaker initial magnetic fields, the total GW amplitudes 
after bounce stay almost zero, because the contribution 
from the magnetic fields cancels with the one from the hydrodynamic counterpart.
 These features can be clearly understood with a careful analysis on the explosion 
dynamics. We point out that the GW signals with the increasing trend, possibly visible 
to the next-generation 
detectors for a Galactic supernova, 
 would be associated with MHD explosions with the explosion energies 
exceeding $10^{51}$ erg.
\end{abstract}

\keywords{supernovae: collapse --- gravitational waves --- 
neutrinos --- hydrodynamics}

\maketitle 
\section{Introduction}

Successful detection of neutrinos from SN1987A paved the way for 
{\it Neutrino Astronomy} \citep{hirata,bionta}, alternative to conventional 
astronomy by electromagnetic waves.
 Core-collapse supernovae are now 
expected to be opening yet another astronomy, {\it Gravitational-Wave Astronomy}.
 Currently long-baseline laser interferometers 
LIGO \citep{firstligonew},
VIRGO$^1$\footnotetext[1]{http://www.ego-gw.it/},
GEO600$^2$\footnotetext[2]{http://geo600.aei.mpg.de/},
TAMA300 \citep{tamanew},
and AIGO$^3$\footnotetext[3]{http://www.gravity.uwa.edu.au/}
with their 
international network of the observatories, 
are beginning to 
take data at sensitivities where astrophysical
events are predicted 
(see, e.g., \citet{hough} for a recent review).  
For these detectors, core-collapse supernovae have been proposed as one of 
the most plausible sources of gravitational waves (GWs) 
(see, e.g., \citet{kota06,ott_rev} for recent reviews).

 Although the explosion mechanism of core-collapse supernovae
 has not been completely clarified yet,
  current multi-dimensional simulations based on refined numerical models
 show several promising scenarios. Among the candidates 
is the neutrino heating mechanism aided by convection and standing accretion shock
 instability (SASI) (e.g., \citet{marek,bruenn,sche04,fog,suwa}), 
the acoustic mechanism \citep{burr06,burrows2}, and the 
magnetohydrodynamic (MHD) mechanism
 (e.g., \citet{arde00,kota04b,kotake05,ober06b,burr07,taki09} and references therein). 
For the former two to be the case, 
the explosion geometry is expected to be unipolar and bipolar, and for 
 the MHD mechanism to be bipolar.

  Since the GW signatures imprint a live information
 of the asphericity at the moment of explosion, they are expected to
 provide us an important hint to solve the supernova mechanism.
So far, most of the theoretical predictions of GWs 
have focused on the bounce signals in the context of rotational core-collapse 
(e.g., \citet{mm,zweg,kotakegw,shibaseki,ott,ott_prl,ott_2007,dimm02,dimmelprl,dimm08,simon}). 
For the bounce signals having a strong and characteristic 
  signature, the iron core must rotate enough rapidly. 
 The waveforms are categorized into the following three types, 
namely types I, II, and III. Type II and III waveforms are 
 shown less likely to appear than type I, because a combination of general 
relativity (GR) and electron capture near core bounce suppresses 
 multiple bounce in the type II waveforms
 \citep{dimmelprl,ott_prl,ott_2007}. In general, a realistic nuclear 
 equation of state (EOS) is 
stiff enough to forbid the type III waveforms. After bounce, asymmetries 
 due to convection \citep{burohey,muyan97,fryersingle,mueller04}, SASI
 \citep{marek_gw,kotake07,kotake09,kotake_ray,murphy}, and g-mode oscillations of 
protoneutron stars \citep{ott_new}, are expected to account for sizable GW signals. 

 In general, detection of these GW signals in the postbounce phase (except 
 for the g-mode oscillation) is far more difficult than the bounce signals, because they 
do not possess a clear signature like bounce signals, but change 
 stochastically with time as a result of chaotically growing convection as
 well as SASI in the non-linear hydrodynamics (\citet{kotake09,kotake11,marek_gw,murphy}).

Rapid rotation, necessary for the strong bounce signals,  
is likely to obtain $\sim$ 1\% of massive star 
population (e.g., \citet{woos_blom}). However this can be really the case for
 progenitors of rapidly rotating metal-poor stars,
 which experience the so-called chemically homogeneous evolution \citep{woos06,yoon}. 
 The high angular momentum of the core as well as a strong precollapse magnetic field
 is preconditioned for the MHD mechanism, 
because the MHD mechanism relies on the extraction of rotational free energy of the 
collapsing core via magnetic fields. The energetic MHD explosions
 are receiving great attention
 recently as a possible relevance to magnetars and collapsars
 (e.g., \citet{harikae_a,harikae_b} for collective references), 
 which are presumably linked to the formation of long-duration gamma-ray bursts (GRBs)
 (e.g., \citet{mesz06}).

 Among the previous studies mentioned above, only a small portion of papers
has been spent on determining the GW signals in the MHD mechanism
 \citep{yama04,kota04a,ober06b,cerd07,shib06,scheid}.  
 This may be because the MHD effects on the dynamics as well as
 their influence over the GW signals can be visible 
 only for cores with precollapse magnetic fields over $B_{0} \gtrsim 10^{12}$ G 
\citep{ober06b,kota04a}. 
Considering that the typical magnetic-field strength of GRB 
progenitors is at most $\sim 10^{11-12}$ G \citep{woos06}, 
this is already an extreme situation. 
Interestingly in a more extremely case of
 $B_0 \sim 10^{13}$ G, a secularly growing feature in the waveforms
was observed \citep{ober06a,shib06,scheid}. 
 Moreover \citet{ober06a} called a waveform as type IV in which 
  quasi-periodic large-scale oscillations of GWs near bounce are
replaced by higher frequency irregular oscillations. 
 Some of these MHD simulations follow adiabatic core-collapse, in which a polytropic 
EOS is employed to mimic supernova microphysics. At this level of approximation, 
the bounce shock generally does not stall and a prompt explosion occurs 
within a few ten milliseconds after bounce. Therefore a main focus in these 
 previous studies has 
been rather limited to the early postbounce phase ($\lesssim$ several 10 ms).
 However, for models with weaker precollapse magnetic fields akin to the current GRB
 progenitors, the prompt shocks stall firstly in the core like a conventional 
 supernova model with more sophisticated neutrino treatment (e.g., \citet{burr07}). 
In such a case, the onset of MHD explosions, depending on the initial 
 rotation rates, can be delayed till $\sim 100$ ms after bounce \citep{burr07,taki09}.
 There remains a room to study GW signatures in such a case, which we hope to 
 study in this work.

 In this study, we choose to take precollapse magnetic fields less than  
 $10^{12}$ G based on a recent GRB-oriented progenitor models.
  By this choice, it generally takes much 
longer time after bounce than the adiabatic MHD models 
to amplify magnetic fields enough strong
 to overwhelm the ram pressure of the accreting matter, leading to the 
 magnetohydrodynamically-driven (MHD, in short) explosions.
 Even if the speed of jets in MHD explosions 
is only mildly relativistic, Newtonian simulations 
 are not numerically stable because the Alfv\'{e}n velocity ($\propto B/\sqrt{\rho}$) 
 could exceed the speed of light unphysically especially when the strongly 
 magnetized jets (i.e., large $B$) propagate to a stellar envelope with decreasing
 density ($\rho$).
 To follow a long-term postbounce evolution numerically stably, 
we perform special relativistic MHD (SRMHD) simulations \citep{taki09}, in which 
 a realistic EOS is employed and the neutrino cooling is taken into account via a 
multiflavor neutrino leakage scheme. 
 Note in our previous study of GWs in magneto-rotational core-collapse 
 \citep{kota04a} that we were unable to study
properties of the GWs long in the postbounce phase because the employed 
 Newtonian simulations quite 
 often crashed especially in the case of strong MHD explosions.
To include GR effects in this study, we follow 
 a prescription in \citet{ober06a} which is reported to 
capture basic features of full GR simulations quite well. 
By changing precollapse magnetic fields as well as initial
 angular momentum distributions parametrically, we compute twelve models.
By doing so, we hope to study the properties of GWs in MHD explosions systematically 
and also address their detectability.

The paper opens up with descriptions of the initial models and numerical methods 
employed in this work (section \ref{sec2}). 
 Formalism for calculating the gravitational waveforms in SRMHD
is summarized in section 
\ref{sec3}. The main results are given in Section \ref{sec4}.  
We summarize our results and discuss their implications in Section \ref{sec5}.

\section{Models and Numerical Methods  \label{sec2}}

\subsection{Initial Models}\label{sec:IM}

We make precollapse models by taking the profiles of density,
internal energy, and electron fraction distribution from
a rotating presupernova model of E25 in \citet{hege00}.
This model has mass of $25M_{\odot}$ at the zero-age main
sequence, however loses the hydrogen envelope and becomes a Wolf-Rayet (WR)
star of 5.45 $M_\odot$ before core-collapse.
Our computational domain involves the whole iron-core of $1.69 M_{\odot}$.
 Note that this model is suggested as a candidate progenitor of 
long-duration GRBs because type Ib/c core-collapse supernovae originated from WR stars
 have a observational association with long-duration GRBs (e.g., \citet{woos_blom}).

Since little is known about the spatial distributions of rotation and
 magnetic fields in evolved massive stars, we add the following profiles 
in a parametric manner to the non-rotating core mentioned above.
For the rotation profile, we assume a cylindrical rotation of 

\begin{equation}
 \Omega(X,Z)
=\Omega_{0}\frac{X_{0}^2}{X^2+X_{0}^2}
\frac{Z_{0}^4}{Z^4+Z_{0}^4}
\label{eq:CR},
\end{equation}
where $\Omega$ is angular velocity and $X$ and $Z$ denotes distance
from the rotational axis and the equatorial plane, respectively.
The parameter $X_{0}$ represents the degree of differential rotation, 
 which we choose to change in the following three ways, 
100km (strongly differential rotation), 500km,
 (moderately differential rotation), and 2000km (uniform rotation), respectively.
The parameter $Z_{0}$ is fixed to 1000km.

Regarding the precollapse magnetic field,
we assume that the magnetic field is nearly uniform and parallel to the rotational axis 
in the core and dipolar outside. This can be 
 modeled by the following effective vector potential,
\begin{equation}
A_r=A_\theta=0,
\end{equation}
\begin{equation}
 A_\phi=\frac{B_0}{2}\frac{r_0^3}{r^3+r_0^3}r\sin\theta,\label{vec_phi}
\end{equation}
where $A_{r,\theta,\phi}$ is vector potential in the $r,\theta,\phi$ 
direction, respectively,  $r$ is radius, $r_0$ is radius of the core, and 
$B_0$ is a model constant  (see \citet{taki04} for detail). 
In this study, $r_0$ is set to $2000$ km which
is approximately the size of the precollapse iron core.

By changing initial angular momentum, degree of differential rotation,  
and the strength 
of magnetic fields, we compute twelve models.
The model parameters are shown in Table \ref{tab:model}.
The models are named 
after this combination,
with the first letters, B12, B11 representing strength of the initial magnetic field,
the second letters, $X1, X5, X20$ indicating the degree of differential rotation
 ($X_0 = 100, 500, 2000$ km, respectively), and the third letter, $\beta=0.1, 1$ 
 showing the rotation parameter $\beta$.
Here $\beta_{\rm}$ represents ratio of 
the rotational energy to the absolute value of the gravitational energy prior 
 to core-collapse.
 The original progenitor of model E25 in \citet{hege00} 
has a uniform rotation profile in the iron core and 
 the initial $\beta$ parameter is $\sim$ 0.15 \%. 
So the initial angular momentum in our models of B11X20$\beta0.1$ and
 B12X20$\beta0.1$ are similar to the original one.
 In a GRB-oriented progenitor of model 35OB \citep{woos06}, 
the precollapse magnetic fields reach to $\sim 10^{11} - 10^{12}$ G and 
 $\beta \sim 0.2 \%$, which is not so different from the chosen parameters 
here.
 Although little is known about the degree of precollapse 
 differential rotation, 
an extremely strong one, for example,
 our model series of X1, should be unrealistic. These models,
 albeit rather academic, are examined in order to see clearly the effects 
of differential rotation.

\begin{table}[h]
\begin{center}
\begin{tabular}{c|ccccccc}
\tableline\tableline
  &    &         &   &{$\beta(\%)$} & &   &   \\
  &    &  0.1\%  &   &              & &1\% &   \\
  &    &         &   & $X_0 (\rm km)$        & &   &   \\
  &       $100{\rm km}$  & $500{\rm km}$  & $2000{\rm km}$ & &
          $100{\rm km}$  & $500{\rm km}$  & $2000{\rm km}$ \\
\tableline
$B_0 : 10^{11}\gauss$  & B11X1$\beta$0.1  &B11X5$\beta$0.1  & B11X20$\beta$0.1 
&  &  B11X1$\beta$1 & B11X5$\beta$1 & B11X20$\beta$1  \\
$~~~~~~10^{12}\gauss$  &  B12X1$\beta$0.1  &B12X5$\beta$0.1  & B12X20$\beta$0.1 
 & &  B12X1$\beta$1 & B12X5$\beta$1 & B12X20$\beta$1  \\
\tableline
\end{tabular}
 \caption{Summary of initial models.
Model names are labeled by the precollapse magnetic fields and rotation.
$\beta$ represents ratio of initial rotational energy to the absolute value of
  the initial gravitational energy.
From left to right in the table, $\Omega_{0}$ in unit of rad/s 
(equation (\ref{eq:CR})) is 24, 2.8, 0.95, 76, 8.9, and 3.0, respectively.
Note that $X_0$ and $B_{0}$ is defined in equation (1) and (\ref{vec_phi}), 
respectively.}
    \label{tab:model}
\end{center}
\end{table}

\subsection{Special Relativistic Magnetohydrodynamics}\label{sec:SRM}

Numerical results in this work are calculated by the SRMHD code developed 
in \citet{taki09}. In the following, we first mention the importance of SR 
and then briefly summarize the numerical schemes.  

The Alfv\'{e}n velocity of MHD jets
propagating into the outer layer of the iron core can be estimated as 
$v_{A} \sim 10^{10} \mathrm{cm/s}~ {(B/10^{13}
\mathrm{G})}{(\rho/ 10^5\gpcmc)}^{-1/2}$,
where $\rho$ and $B$ are the typical density and the magnetic field
 near along the rotational axis. 
It can be readily inferred that the Alfv\'{e}n velocity 
could exceed the speed of light unphysically in Newtonian simulations. 
 SR corrections are also helpful to capture correctly the dynamics of 
infalling material in the vicinity of the protoneutron star, 
because their free-fall velocities and rotational velocities become close to the 
speed of light. Such conditions are quite ubiquitous in MHD explosions.
 Even if the propagation speeds of the jets are only mildly relativistic, 
we have learned that (at least) SR treatments are quite important 
for keeping the stable numerical calculations in good accuracy over a long-term 
 postbounce evolution (e.g., \citet{harikae_a}).

The MHD part of our code is based on the formalism of \citet{devi03}.
The state of the relativistic fluid element at each point in the space time
is described by its density, $\rho$; specific energy, $e$; velocity,
$v^i$; and pressure, $p$.
And the magnetic field in the laboratory frame is described by
the four-vector $\sqrt{4\pi}b^{\mu}={^*F}^{\mu\nu}U_{\nu}$, where $^*F^{\mu\nu}$ is the
dual of the electro-magnetic field strength tensor and $U_{\nu}$ is the
four-velocity. The basic equations of the SRMHD code are written as,

\begin{eqnarray}
\pdel{t}{D}
+\frac{1}{\sqrt{\gamma}}\udel{i}{\sqrt{\gamma}Dv^i} &=&0 \label{eq:mass_consv}\\
\pdel{t}{E}
+\frac{1}{\sqrt{\gamma}}\udel{i}{\sqrt{\gamma}Ev^i}
&=&-p\pdel{t}{W}
-\frac{p}{\sqrt{\gamma}}\udel{i}{\sqrt{\gamma}W v^i}
-{\cal L}_\nu \label{eq:ene_consv}\\
\pdel{t}{(S_i-b^tb_i)}
+\frac{1}{\sqrt{\gamma}}
\udel{j}{\sqrt{\gamma}\left(S_i v^j-b_ib^j\right)}
&=&
-\frac{1}{2}
\left(
\rho h \left(Wv_k\right)^2
- \left(b_k\right)^2
\right)\udel{i}{\gamma^{kk}}\nonumber\\
& &
-\left(\rho h W^2- {b^t}^2\right) \udel{i}{\Phi_\mathrm{tot}}\nonumber\\
& &
-\udel{i}{\left(p+\frac{|b|^2}{2}\right)}\label{eq:mom_consv}\\
\pdel{t}{(Wb^i - Wb^tv^i)}
+\udel{j}{\left(Wv^jb^i-Wv^ib^j\right)}
&=&0\label{eq:induction}\\
\tdel{k}{{\udel{k}{\Phi}}}&=&4\pi \biggl[
\rho h (W^2+(v_k)^2) 
+ 2\left( p+\frac{\left|b\right|^2}{2}\right) \nonumber \\
&&
-\left((b^{0})^2+(b_{k})^2 \right) \biggl]
\label{eq:poisson}
\end{eqnarray}
where $W=\frac{1}{\sqrt{1-v^kv_k}}$, $D=\rho W$, $E=e W$ and 
$S_i=\rho hW^2v_i$ are 
 the Lorentz boost factor, auxiliary variables correspond to density,
energy, and momentum, respectively. 
 All of them are defined in the laboratory frame. 
Equations (\ref{eq:mass_consv},\ref{eq:ene_consv},\ref{eq:mom_consv})
represents the mass, energy, and momentum conservation, respectively.
In equation (\ref{eq:mom_consv}), note that the relativistic enthalpy,
$h=(1+e/\rho+p/\rho+\left|b\right|^2/\rho)$
 includes the magnetic energy.
Equation (\ref{eq:induction}) is induction equation in SR. 
In solving the equation, the method of characteristics
is implemented to propagate accurately all modes of MHD waves 
(see \citet{taki09} for more detail).
Equation (\ref{eq:poisson}) is Poisson equation for 
 the gravitational potential of $\Phi$, which is solved by the modified incomplete
 Cholesky conjugate gradient method. For an approximate treatment 
of general relativistic (GR) gravity,  $\Phi_{\rm tot}$ in equation (\ref{eq:mom_consv}) 
includes a GR correction to $\Phi$ as in \citet{buras06}.
We employ a realistic equation of state 
based on the relativistic mean field theory \citep{shen98}.

We approximate the neutrino cooling by a neutrino multiflavor leakage
scheme \citep{epst81,ross03}, in which three neutrino flavors:
electron neutrino($\nu_{e}$), electron antineutrino($\bar{\nu}_{e}$), and
 heavy-lepton neutrinos($\nu_{\mu}$, $\bar{\nu}_{\mu}$,
$\nu_{\tau}$, $\bar{\nu}_{\tau}$, collectively referred to
as $\nu_{X}$), are taken into account. 
The implemented neutrino reactions 
 are electron capture on
proton and free nuclei; positron capture on neutron; photo-, pair, plasma
processes \citep{full85,taka78,itoh89,itoh90}. 
 A transport equation for the lepton fractions (namely 
$Y_e-Y_{e^{+}},Y_{\nu_e}, Y_{\bar{\nu}_e}$ and $Y_{{\nu_{X}}},$ 
are solved in an operator-splitting
manner (see equation (7) in \citet{taki09} for more detail).
 $\cal{L}_{\nu}$ in equation (\ref{eq:ene_consv}) represents the 
neutrino cooling rate summed over all the reactions, 
which can be also estimated by the leakage scheme
 (see \citet{epst81,ross03,kota03a} for more detail).

In our two dimensional simulations, the spherical coordinates are
 employed with 300($r$) $\times$ 60($\theta$) grid points to cover the
computational domain assuming axial and equatorial symmetry.
The radial grid is nonuniform,
extending from $0$ to $4000 \km$ with finer
grid near the center. The finest grid for the radial direction is set to $1 \km$.
The polar grid uniformly covers from $\theta=0$ to ${\pi}/{2}$. 
The finest grid for the polar direction is $25$ m.
The numerical tests and convergence with 
this choice of the numerical grid points are given in section 5 of \citet{taki09}.

To measure the strength of explosion, we define the explosion energy as follows,
\begin{equation}
E_{\mathrm{exp}} = \int_{\mathrm{D}} dV~e_{\mathrm{local}}
=
\int_{\mathrm{D}} dV \left(e_{\mathrm{kin}}+e_{\mathrm{int}}+e_{\mathrm{mag}}+e_{\mathrm{grav}}\right),\label{eq:expene}
\end{equation}
here $e_{\mathrm{local}}$ is the sum of $e_{\mathrm{kin}}$, $e_{\mathrm{int}}$, $e_{\mathrm{mag}}$ and
$e_{\mathrm{grav}}$ with being kinetic, internal, magnetic, and gravitational 
energy, respectively defined as 
\begin{eqnarray}
 e_{\mathrm{kin}}&=&\rho W \left(W-1\right),\\
 e_{\mathrm{int}}&=& e W^2+p\left(W^2-1\right),\\
 e_{\mathrm{mag}}&=&  |b|^2\left(1-\frac{1}{2W^2}\right)
-\frac{{b^0}^2}{2W^2},\\
 e_{\mathrm{grav}}&=&-\rho h W^2 \Phi,
\end{eqnarray}
 and $\mathrm{D}$ in equation (9) represents
the domain where the local energy ($e_{\mathrm{local}}$) 
is positive, indicating that the matter
is gravitationally unbound. The explosion energy is evaluated when
the MHD jets pass through $1000$ km along the polar direction. 

\section{Formulae for Gravitational Waves in SRMHD \label{sec3}}
 To extract the gravitational waveforms in MHD explosions, 
we present the stress formula in SRMHD for later
 convenience. As shown below, this can be done straightforwardly 
by extending the Newtonian MHD formulation presented in \citet{kota04a}. 

From the Einstein equation, one obtains the following formula as a 
primary expression for the leading part of the gravitational quadrupole field 
emitted by the motion of a fluid in the post-Newtonian approximation 
 (e.g., \citet{mm,finn,blanchet}),
\begin{eqnarray}
h_{ij}^{TT}({\mbox {\boldmath  $X$}},t) &=& \frac{4G}{c^4 R}P_{ijkl}
({\mbox {\boldmath  $N$}}) \int d^3 x~ T_{kl},
\label{1st}
\end{eqnarray} 
where $G$ and $c$ are the gravitational constant and the velocity of light, 
respectively, $T_{kl}$ is the energy momentum tensor of the source, 
$R \equiv (\delta_{ij} X^i X^j)^{1/2} = |{\mbox {\boldmath  $X$}}|$ is the distance 
between the observer and the source.  $P_{ijkl}$, with 
 ${\mbox {\boldmath  $N$}} = {\mbox {\boldmath  $X$}}/R$ 
denotes the transverse-traceless (TT) projection operator 
 onto the plane orthogonal to the outgoing wave direction ${\mbox {\boldmath  $N$}}$
 (e.g., \citet{mm}) .

 $T_{ij}$ consists of the three parts, namely of perfect fluid, electromagnetic field, and gravitational potential as follows,
\begin{eqnarray}
T_{ij}&=& {T_{ij}}_{\rm(hyd)} + {T_{ij}}_{\rm(mag)} + {T_{ij}}_{\rm(grav)}.
\label{2nd}
\end{eqnarray}
The first term which we refer as the hydrodynamic part is explicitly written as, 
\begin{eqnarray}
{T_{ij}}_{\rm(hyd)} &=& \rho_{*} W^2 v_i v_j + p \delta_{ij},
\label{3rd}
\end{eqnarray}
 where $\rho_{*}$  
  is effective density defined as, 
\begin{equation}
\rho_{*}= \rho + \frac{e + p + |b|^2}{c^2}.
\end{equation}
The second term in equation (15) represents the contribution from magnetic fields as,
\begin{eqnarray}
{T_{ij}}_{\rm(mag)} &=& -b_i b_j.
\end{eqnarray}
And the last term in equation (15) is the contribution from the gravitational potential, 
\begin{eqnarray}
{T_{ij}}_{\rm(grav)} &=& \frac{1}{4\pi G}\Biggl(
\Phi_{,i}\Phi_{,j} - \frac{1}{2}\delta_{ij}\Phi_{,m}\Phi^{,m}\Biggr)
\end{eqnarray}
 where $\Phi$ corresponds to the self-gravity in equation (\ref{eq:poisson}). 

In our axisymmetric case, there remains only one non-vanishing quadrupole term
 in the metric perturbation, namely $\ell = 2, m = 0$ in terms of the pure-spin tensor harmonics, as
\begin{equation}
h_{ij}^{TT}({\mbox {\boldmath  $X$}},t) 
\stackrel{\ell = 2, m = 0}{=} \frac{1}{R}A^{E2}_{20}\left(t - 
\frac{R}{c}\right)T^{E2,20}_{ij}(\theta,\phi),
\label{htt}
\end{equation}
 where $T^{E2,20}_{ij}(\theta,\phi)$ is 
\begin{equation}
T^{E2,20}_{ij}(\theta,\phi) = \frac{1}{8}\sqrt{\frac{15}{\pi}}\sin^2 \theta,
\label{Tij}
\end{equation}
(e.g., \cite{thorne}).
 The projection operator in equation (\ref{1st}) acts on $T_{ij}$ as,
\begin{equation}
P_{ijkl} T^{kl} = 2 T^{zz} - T^{xx} - T^{yy}.
\end{equation}
Transforming equation (\ref{1st}) to the spherical coordinates,
and expressing $b_i$ and $v_i$ in terms of unit vectors in the 
 $r$, $\theta$, $\phi$ direction, 
we obtain for ${A_{20}^{\rm{E} 2}}$ the expression,
\begin{equation}
 {A_{20}^{\rm{E} 2}} =  {A_{20}^{\rm{E} 2}}_{\rm (hyd)} + 
 {A_{20}^{\rm{E} 2}}_{\rm (mag)} + {A_{20}^{\rm{E} 2}}_{\rm (grav)},
\label{A20}
\end{equation}
where 
\begin{eqnarray}
 {A_{20}^{\rm{E} 2}}_{\rm(hyd)} &=& \frac{G}{c^4} \frac{32 \pi^{3/2}}{\sqrt{15 }} 
\int_{0}^{1}d\mu 
\int_{0}^{\infty}  r^2  \,dr  {f_{20}^{\rm{E} 2}}_{\rm(hyd)}, \\
{f_{20}^{\rm{E} 2}}_{\rm(hyd)} 
&=& \rho_{*}W^2( {v_r}^2 ( 3 \mu^2 -1) + {v_{\theta}}^2 ( 2 - 3 \mu^2)
 - {v_{\phi}}^{2} - 6 v_{r} v_{\theta} \,\mu \sqrt{1-\mu^2});
\label{quad}
\end{eqnarray}
\begin{eqnarray}
 {A_{20}^{\rm{E} 2}}_{\rm(grav)} &=& \frac{G}{c^4} \frac{32 \pi^{3/2}}{\sqrt{15 }} 
\int_{0}^{1}d\mu  \int_{0}^{\infty}  r^2  \,dr
 {f_{20}^{\rm{E} 2}}_{\rm(grav)} ,\\
{f_{20}^{\rm{E} 2}}_{\rm(grav)}
&=&
 \left[\rho h (W^2+(v_k/c)^2) + 
 \frac{2}{c^2}\left( p+\frac{\left|b\right|^2}{2}\right)
-\frac{1}{c^2}\left((b^{0})^2+(b_{k})^2 \right)\right]\nonumber \\
& & \times \left[- r \partial_{r} \Phi (3 \mu^2 -1) + 3 \partial_{\theta} \Phi \,\mu
\sqrt{1-\mu^2}\right];
\label{grav}
\end{eqnarray}
and 
\begin{eqnarray}
 {A_{20}^{\rm{E} 2}}_{\rm(mag)} &=& - \frac{G}{c^4} \frac{32 \pi^{3/2}}{\sqrt{15 }} 
\int_{0}^{1}d\mu  \int_{0}^{\infty}  r^2  \,dr {f_{20}^{\rm{E} 2}}_{\rm(mag)}, \\ 
 {f_{20}^{\rm{E} 2}}_{\rm(mag)}
&=&
 [{b_r}^2 ( 3 \mu^2 -1) + {b_{\theta}}^2 ( 2 - 3 \mu^2)
 - {b_{\phi}}^{2} - 6  b_{r} b_{\theta} \mu \sqrt{1-\mu^2}];
\label{mag}
\end{eqnarray}
 where $\mu = \cos \theta$. For later convenience,  we write the total 
 GW amplitude as,
\begin{equation}
h^{\rm TT} = h^{\rm TT}_{\rm(hyd)} + h^{\rm TT}_{\rm(mag)} + h^{TT}_{\rm(grav)}, 
\label{total}
\end{equation}
where the quantities of the right hand are defined by combining
equations (\ref{htt}) and (\ref{A20}) with equations
 (\ref{quad}), (\ref{grav}), and (\ref{mag}). 
 By dropping $O(v/c)$ terms, 
the above formulae reduce to the conventional Newtonian stress formula 
(e.g., \cite{mm}).
 In the following computations, the observer is assumed to be located in 
the equatorial plane ($\theta = \pi/2$ in equation (\ref{Tij})), and that the source 
is located near at our galactic center ($R = 10~\rm{kpc}$).

\clearpage

\section{Results}\label{sec4}
 The gravitational waveforms obtained in this work can be categorized into 
two, which we call increasing type or cancellation type just for convenience.
  Note that the latter type does not mean a new waveform as will be explained 
 later in this section. Regarding the former one, such a waveform was presented  
 in previous literature \citep{ober06a,shib06,scheid}, however 
 their properties have not been clearly understood yet.
In section \ref{A}, we first overview their characters, which are peculiar in
 the case of MHD explosions. In section \ref{B}, we move on 
 to analyze their properties by carefully comparing each contribution
 in equation (\ref{total}) to the total GW amplitudes. Then in section \ref{C},
 we perform the spectra analysis and discuss their 
 detectability.

\subsection{Properties of Waveforms in the MHD Exploding Models}\label{A}

\begin{figure}[htbp]
    \centering
    \includegraphics[width=.49\linewidth]{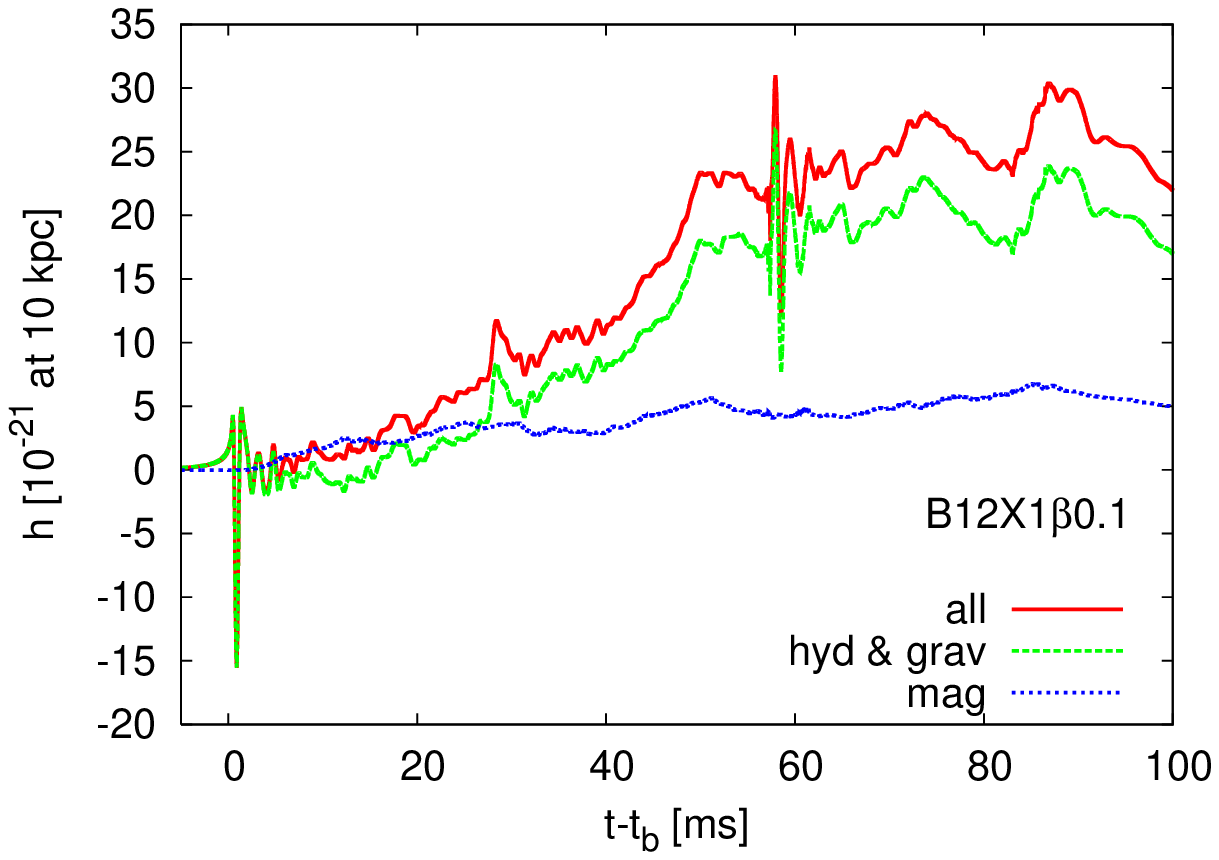}
    \includegraphics[width=.49\linewidth]{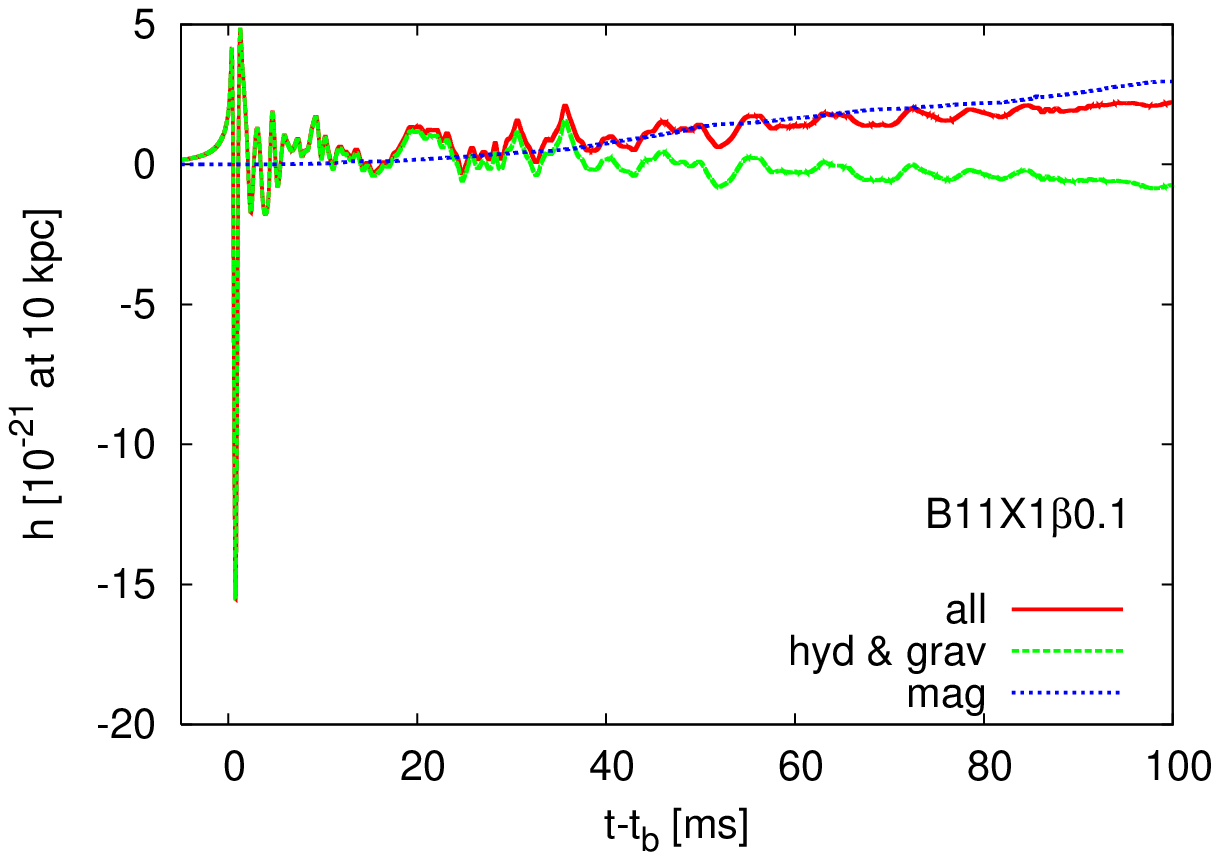}\\
    \includegraphics[width=.49\linewidth]{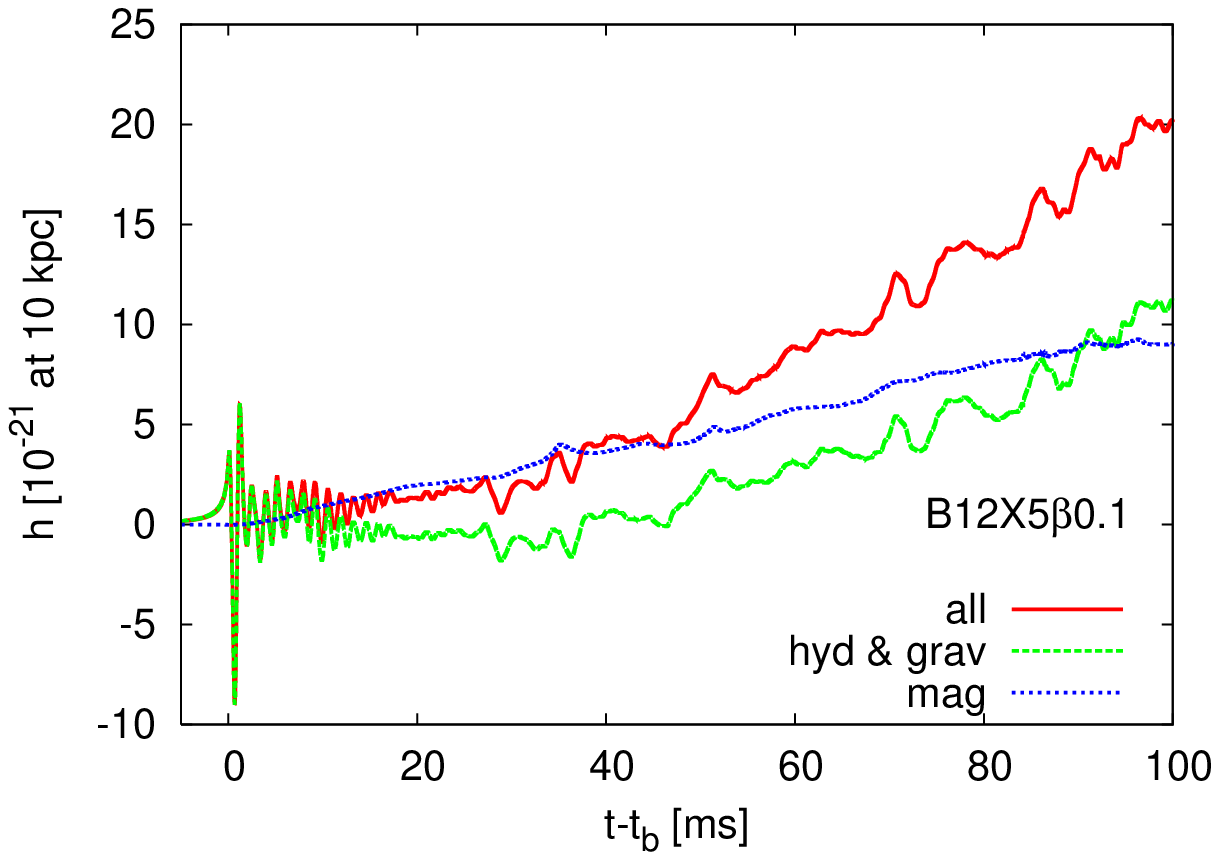}
    \includegraphics[width=.49\linewidth]{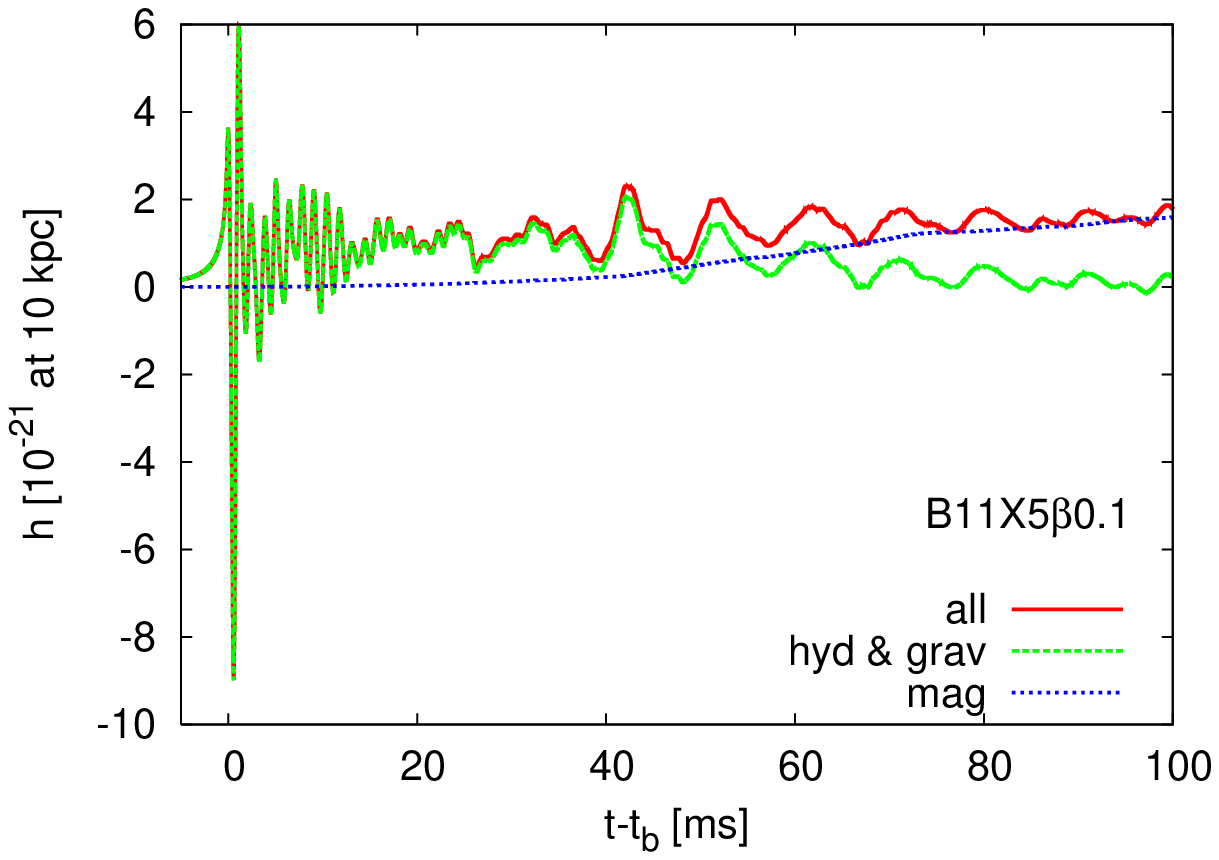}\\
    \includegraphics[width=.49\linewidth]{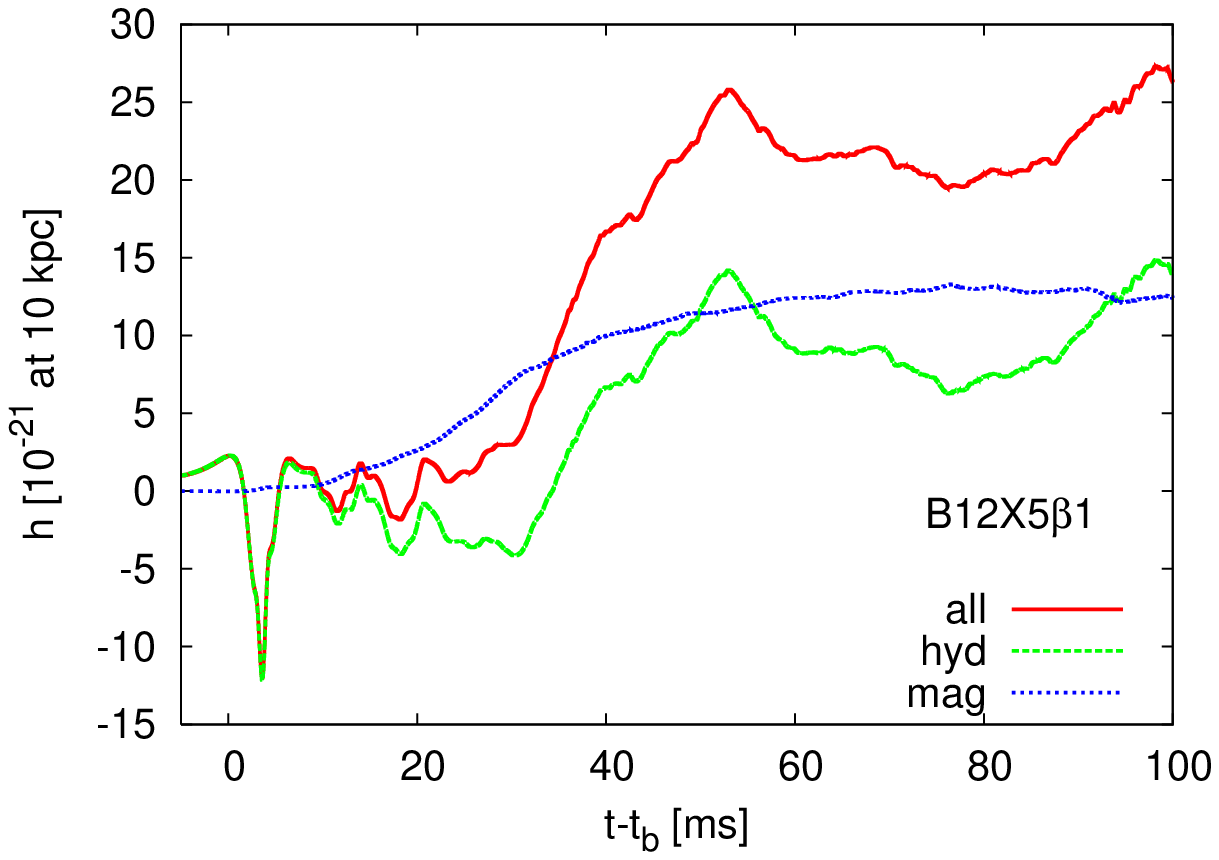}
    \includegraphics[width=.49\linewidth]{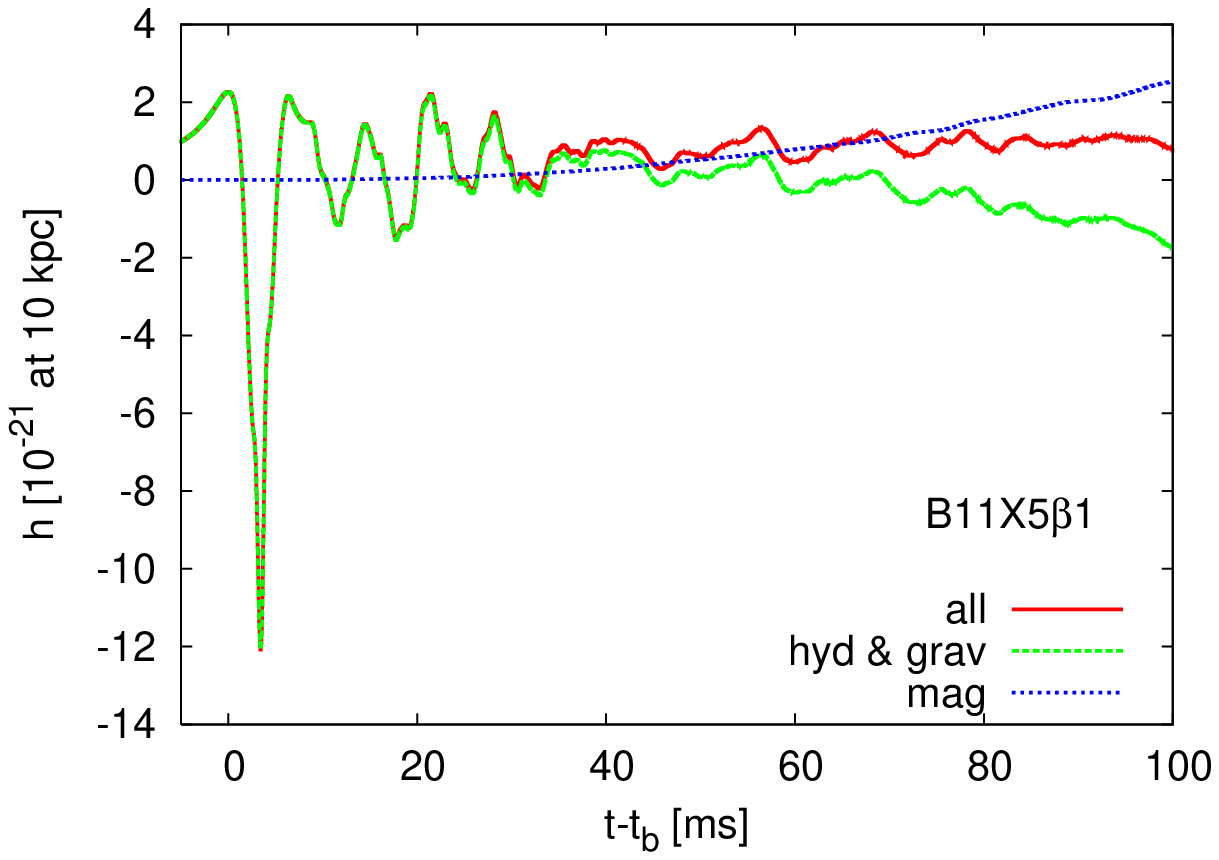}\\
    \caption{Gravitational waveforms with the increasing (left) or the cancellation 
trend (right) (see text for more detail).  At the right bottom in each panel, 
the model names are given such as B12X1$\beta0.1$ (top left, for example). 
  The total wave amplitudes are shown by the red line, while the contribution
 from the magnetic fields and from the sum of the hydrodynamic and gravitational parts
 are shown by blue and green lines, respectively (e.g.,  equation (\ref{mag}) and 
 equations (\ref{quad},\ref{grav})).}
    \label{fig:hakei}
\end{figure}

Figure \ref{fig:hakei} shows examples of the two categories, which we call as 
the increasing (left panels) or cancellation type (right panels), respectively. 
In the increasing type, the total wave amplitudes (red line) have a monotonically
 increase trend after bounce ($t - t_b =0$ in the figures).
 While in the cancellation type 
(right panels), the total amplitudes after bounce stay almost zero. This is 
because the contribution from the magnetic fields (blue line, equation (\ref{mag}))
 cancels with the one from the sum of the hydrodynamic and gravitational parts
 (green line, equations (\ref{quad},\ref{grav})). Regardless of the difference in the 
two types, it is common that the magnetic contribution (blue line) increases almost 
 monotonically with time. Not surprisingly, the bounce GW signals 
(($t - t_b \lesssim 20$ ms) are categorized into the so-called type I or II waveforms. 
 Note here that the MHD simulations are terminated at around 
 100 ms after bounce for all the computed models.
 This is simply because the GW amplitudes in a more later phase 
 decrease because the MHD shock comes out of the computational domain and 
 the enclosed mass in the domain becomes smaller.
 
 Table \ref{table1} depicts a classification of the computed models,
 in which "C" and "I" indicates the cancellation and increasing type,
 respectively. ${\rm I}^{*}$ in the table indicates the mixture of the two types, 
which we call as intermediate type. The table shows that 
 the bifurcation of the two types is predominantly determined by the precollapse 
magnetic fields, so that the models with stronger magnetic fields ($B_0 ~ 10^{12}$ G)
are basically classified to the increasing type. 
 For models colored by orange, which are slow rotator
 with uniform rotation in our models,
 the field amplification works less efficiently
 than for models with stronger 
differential rotation (such as $X_0 = 100, 500$ km) (see also \citet{pablo}).
 This suppresses the 
increase in the wave amplitudes due to magnetic fields, which gives 
rise to the intermediate state between the two types.

 Table \ref{table3} is a summary showing intervals measured from the stall of the 
bounce shock (top) and from core bounce (bottom)
till the MHD-driven revival of the stalled bounce shock. 
As already mentioned in \citet{taki09},
  generation of the postbounce MHD jets proceeds in the following two ways.
 One is launched relatively promptly after the stall of the bounce shock,
 typically earlier than $\sim$ 30 ms (see models colored by red in Table 3)
 and another is launched rather later after bounce 
($\gtrsim 30$ ms) (models colored by green in Table 3).
 In this sense, our computed models could be roughly categorized into two, namely
   promptly MHD explosion (colored by red in Table 3) or  delayed MHD explosion 
(colored by green in Table 3), respectively.
 This simply reflects that it takes longer time for
 the weakly magnetized models to amplify the magnetic pressure behind 
 the stalled shock enough strong
 to overwhelm the ram pressure of accreting matter. 
 By comparing Table \ref{table1} to \ref{table3}, the two characters in the waveforms 
have a rough correlation with the difference of the explosion dynamics.

 Table \ref{table2} shows a summary regarding the
 explosion energy (e.g., equation (\ref{eq:expene})). 
Comparing Table \ref{table1} to \ref{table2},
  the explosion energy for the increasing type (models colored 
 by yellow in Table 2 and 4) is higher compared to the cancellation type 
(models colored by 
 light blue).
 Given the rotation rate of $\beta = 0.1\%$,  the explosion energy 
is smallest for the intermediate type (models colored by orange) due to
  the insufficient field amplification as mentioned above.
 It can be also shown that the explosion energies 
 for the models with the increasing trend generally exceed $10^{51}$ erg
 (models colored by yellow).

Having summarized the waveform classification together with the explosion 
dynamics, we move on to look more in detail 
what makes the difference between the two types in
 the next section.

\begin{table}[htbp]
    \centering
    \includegraphics[width=.8\linewidth]{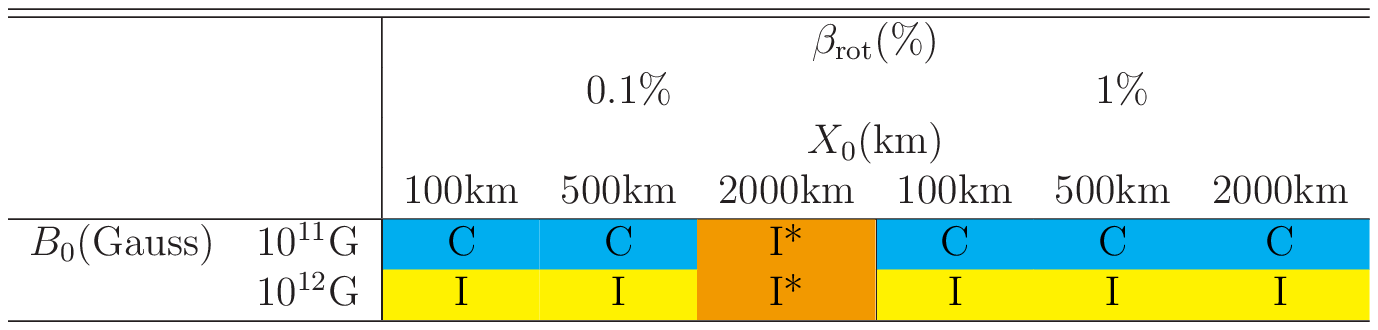}
    \caption{Same as Table \ref{tab:model} but for the classification of the computed 
models. "C" and "I" indicates the cancellation and increasing type, respectively, 
while ${\rm I}^{*}$ indicates the 
mixture of the two types, which we refer to as intermediate type.}
    \label{table1}
\end{table}

\begin{table}[htbp]
    \centering
    \includegraphics[width=.8\linewidth]{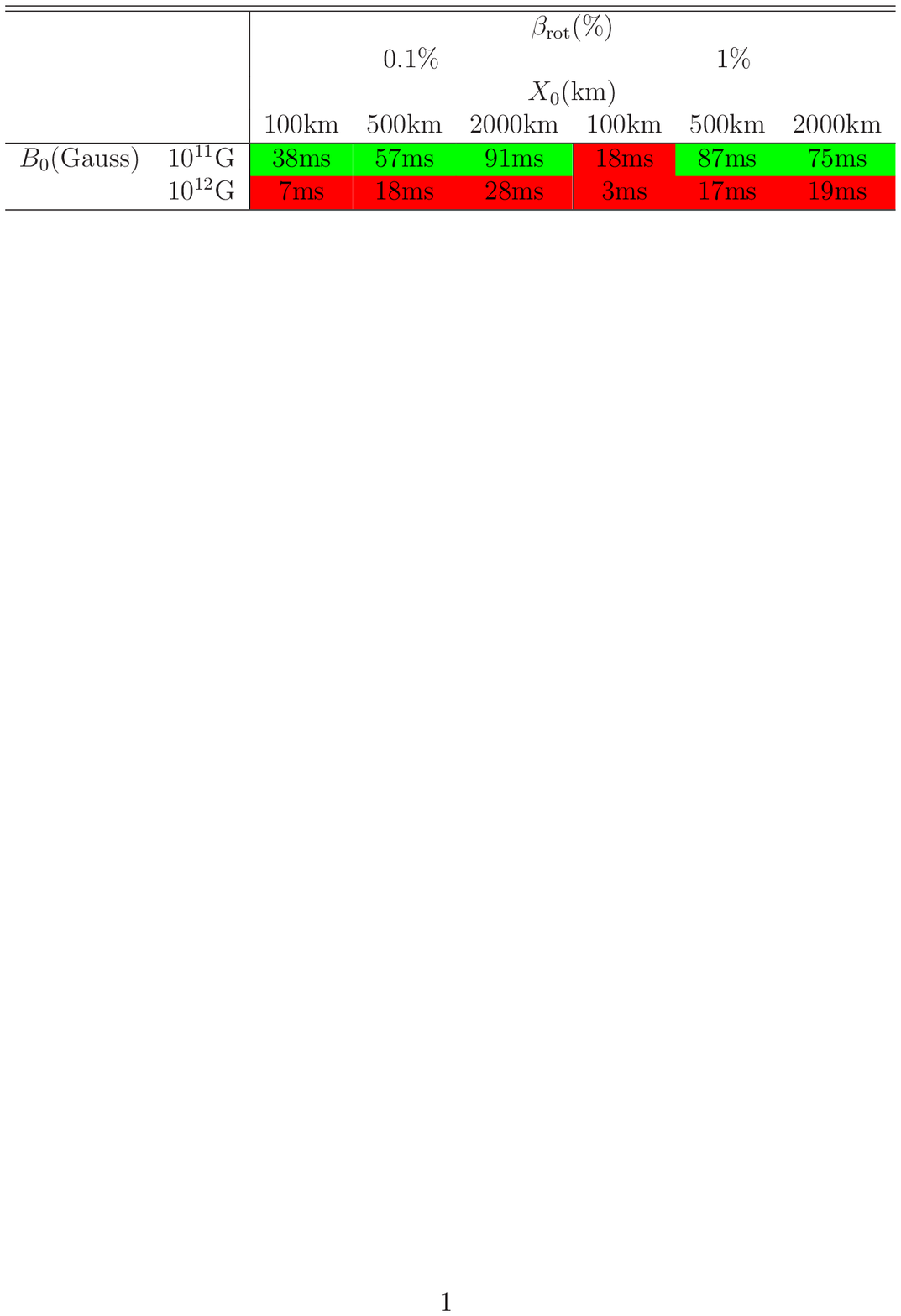}
     \includegraphics[width=.8\linewidth]{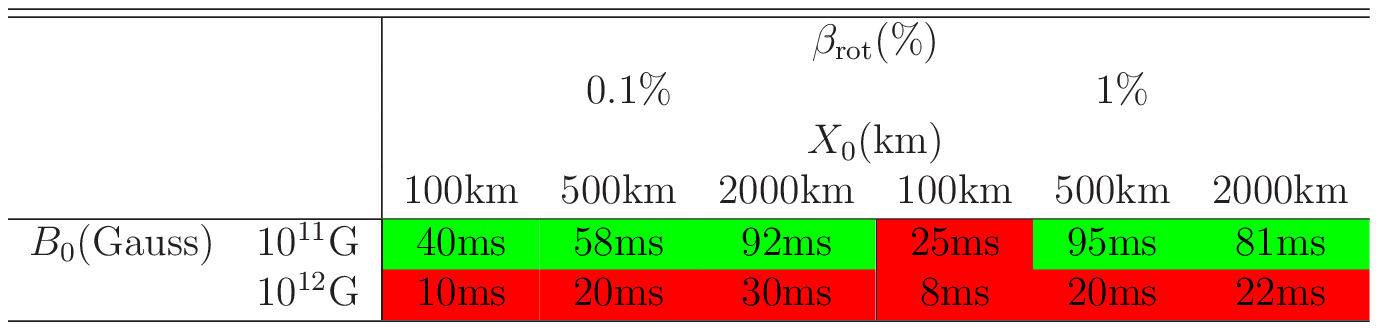}
   \caption{Same as Table \ref{tab:model} but for the interval measured from the 
stall of the bounce shock to the MHD-driven revival of the stalled shock (top panel) 
 and the one measured from core bounce (bottom panel). The computed models 
are classified whether the 
 launch of the MHD jets occurs relatively promptly after bounce
 (models colored by red, with the intervals being shorter than $\sim$ 30 ms typically)
 or rather later (models colored by green), which we refer to as 
promptly or delayed MHD explosion for convenience in this work 
(see text for more detail). Note that these timescales are estimated 
 just by looking at the velocity evolutions along the polar axis.}
    \label{table3}
\end{table}

\begin{table}[htbp]
    \centering
    \includegraphics[width=.8\linewidth]{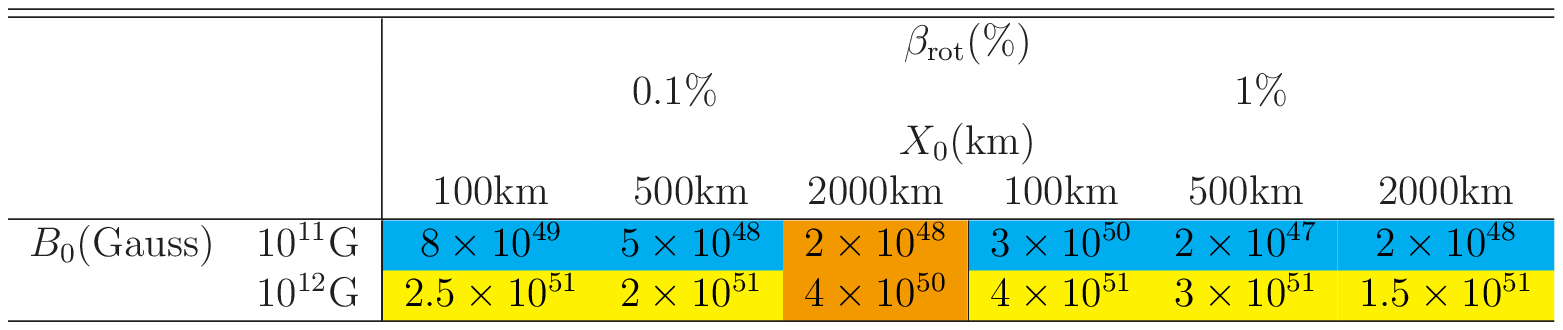}
    \caption{Same as Table \ref{tab:model} but for the explosion energy 
(defined in equation (9)). Comparing with Table \ref{table1} to \ref{table2},
 the explosion energies for the models with the increasing trend generally exceed 
$10^{51}$ erg (models colored by yellow).}
    \label{table2}
\end{table}

\clearpage
\subsection{Analysis of  Waveforms}\label{B}
\subsubsection{Increasing type}
\begin{figure}[htbp]
    \centering
    \includegraphics[width=.44\linewidth]{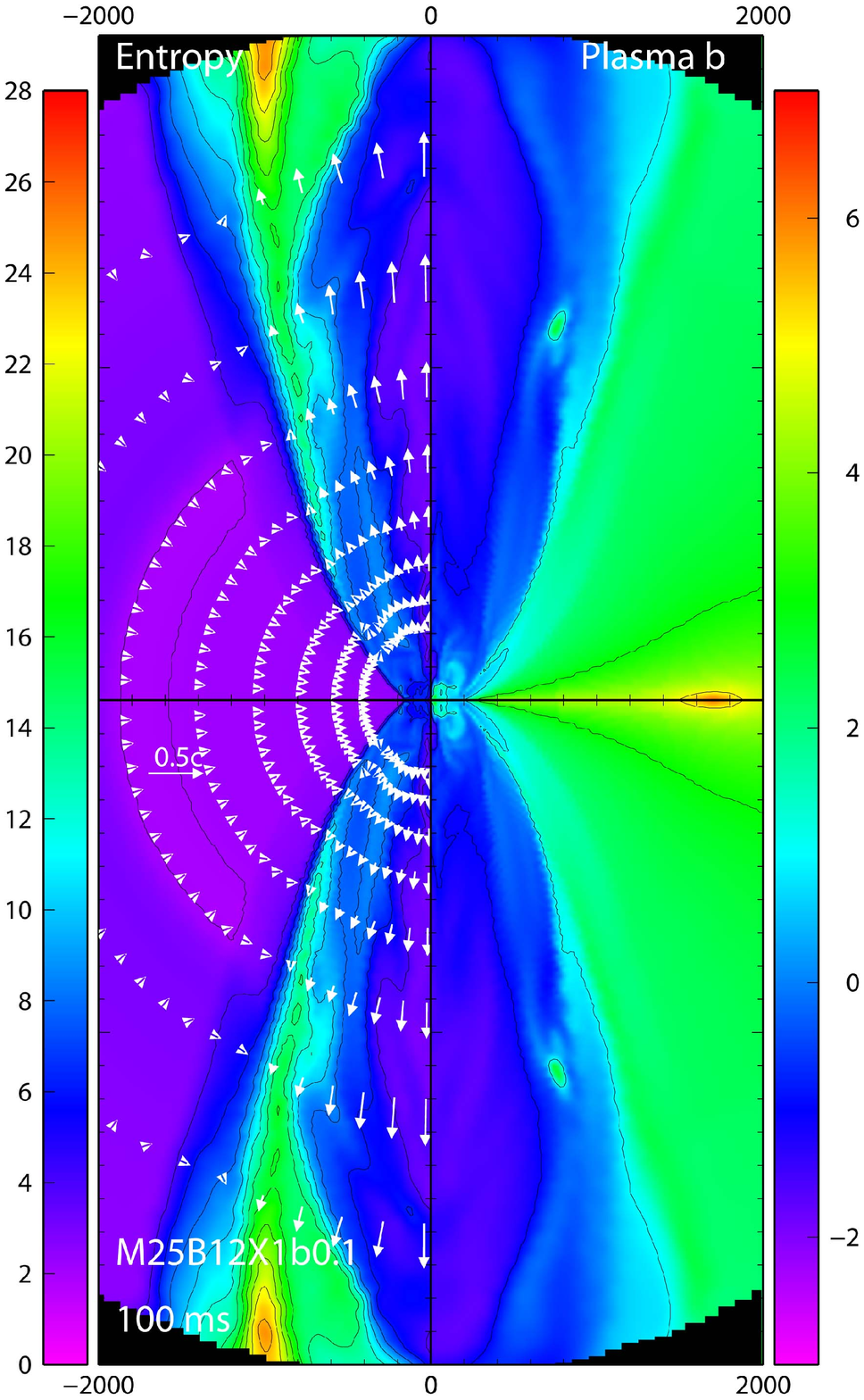}
    \includegraphics[width=.44\linewidth]{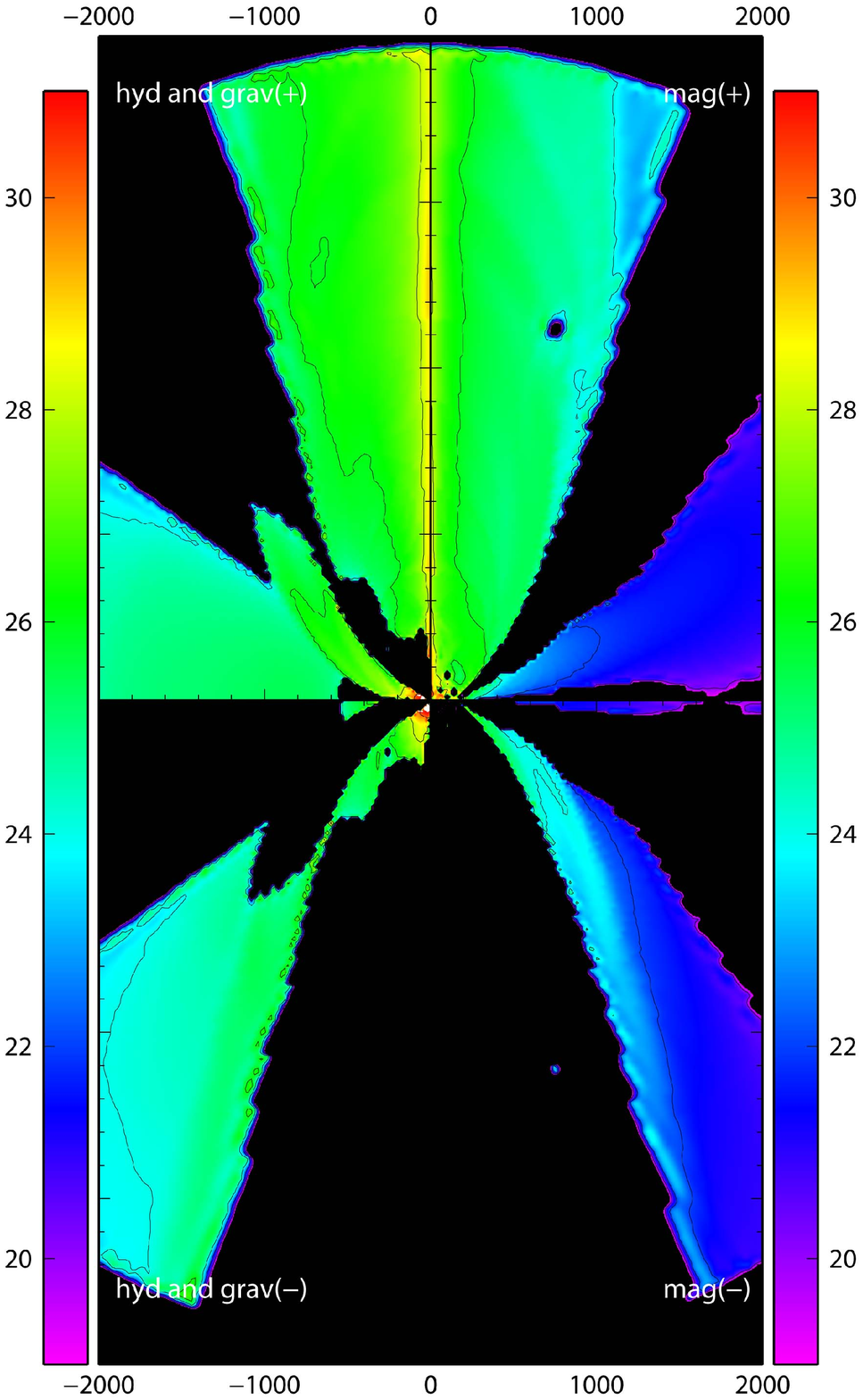}
    \caption{
 Left panel shows distributions of entropy [$k_B$/baryon] (left) and
 logarithm of plasma $\beta$ 
(right) for model B12X1$\beta$0.1 at 100ms after bounce. 
The white arrows in the left-hand side show the velocity fields, 
which are normalized by the scale in the middle left edge ($0.5 c$).
Right panel shows the sum of the hydrodynamic and gravitational parts 
(indicated by ``hyd and grav'' in the left-hand side) and the magnetic part 
 (indicated by "mag" in the right-hand side), respectively. The top and bottom 
panels represent the positive and negative contribution (indicated by (+) or (-))
 to ${A_{20}^{\rm{E} 2}}$,
 respectively (see text for more detail).  The side length of each plot is 
4000(km)x8000(km). }
 \label{fig:inc_cont}
\end{figure}

By taking model B12X1$\beta$0.1 as a reference, we first focus on the
 increase-type waveform.
The left panel of Figure \ref{fig:inc_cont} depicts distributions of entropy
 (left-half) and plasma $\beta$ (right-half, the ratio of matter to magnetic pressure)
 at 100 ms after bounce. 
It can be seen that the outgoing jets indicated 
 by the velocity fields (arrows in the left-half) are driven by the 
 magnetic pressure behind the shock (see bluish region (i.e., low plasma $\beta$) 
in the right-half panel).

The right panel of Figure \ref{fig:inc_cont} shows contributions
 to the total GW amplitudes (equation (27)),
 in which the left-hand-side panels are for the sum of the hydrodynamic and gravitational part,
 namely $\log\left(\pm\left[{f_{20}^{\rm{E} 2}}_{\rm (hyd)} + {f_{20}^{\rm{E} 2}}_{\rm (grav)}\right]\right)$ 
(left top($+$)/bottom($-$)(equations (\ref{quad},\ref{grav})), and the right-hand-side 
panels are for the magnetic part, namely $\log\left(\pm{f_{20}^{\rm{E} 2}}_{\rm (mag)}\right)$ 
(right top($+$)/bottom($-$)) (e.g., equation (\ref{mag})).
 By comparing the top two panels in Figure 2,
 it can be seen that the positive contribution is  
 overlapped with the regions where the MHD outflows exist.
The major positive contribution is from the kinetic term of the MHD outflows 
 with large radial velocities (e.g., $ + \rho_{*}W^2 {v_r}^2$  in equation (\ref{quad})).
 The magnetic part also contributes to the positive trend (see top right-half
  in the right panel (labeled by mag(+))). This comes 
from the toroidal magnetic fields (e.g., $ + {b_{\phi}}^{2}$ in equation
 (\ref{mag})), which dominantly contribute to drive MHD explosions. 
The magnetic contribution was already mentioned in \citet{kota04a}. 
This study furthermore adds that the kinetic energy of MHD outflows 
 more importantly contributes to the positive trend.

\begin{figure}[htbp]
    \centering
    \includegraphics[width=.6\linewidth]{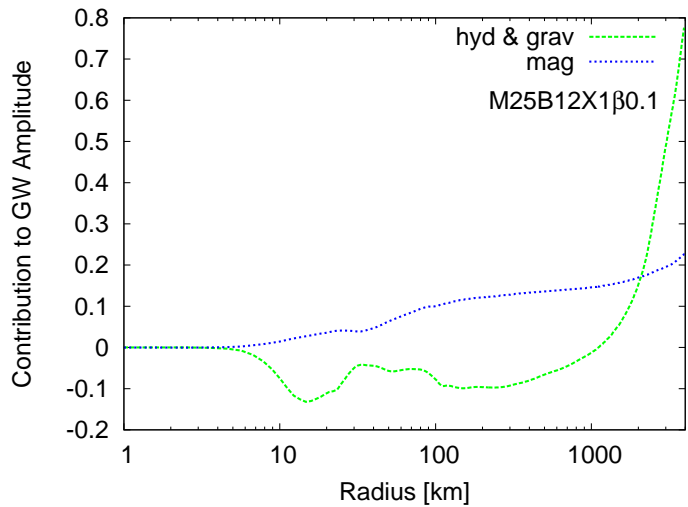}
    \caption{Normalized cumulative contribution of each term in ${A_{20}^{\rm{E} 2}}$ as 
a function of radius for model B12X1$\beta$0.1. This is estimated by the volume 
integral of ${A_{20}^{\rm{E} 2}}$ within sphere of a given radius.}
    \label{fig:inc_rad}
\end{figure}

\begin{figure}
  \begin{center}
    \begin{tabular}{c}
      \resizebox{66mm}{!}{\includegraphics{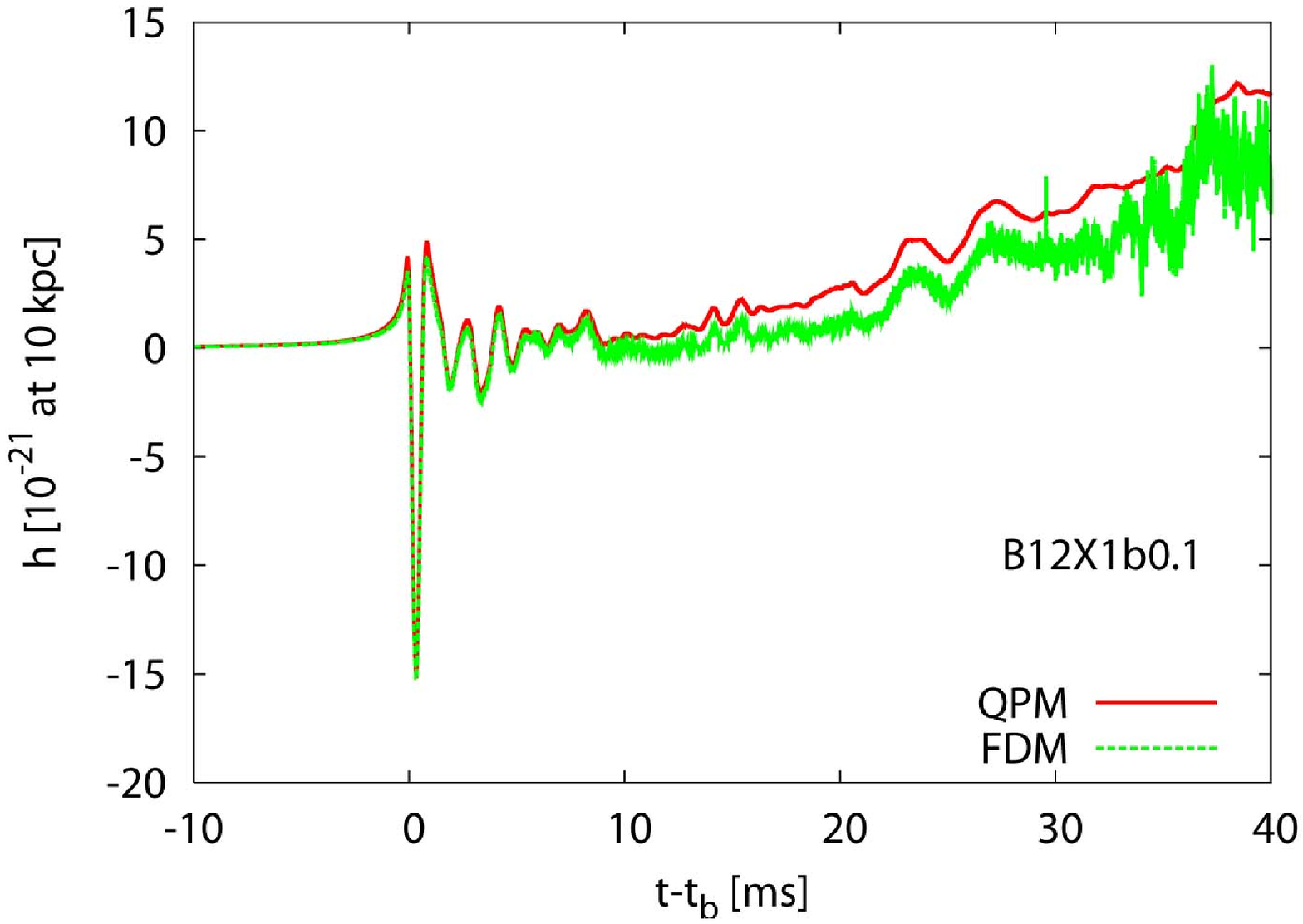}}
    \end{tabular}
   \caption{Gravitational waveform extracted either by the stress formula (indicated by QPM, red line) or by the 
 first-moment of momentum-density formalism (FDM, green line) 
for model B12X1$\beta0.1$), respectively.}
  \label{fig15}
  \end{center}
\end{figure}

Figure \ref{fig:inc_rad} shows a normalized cumulative contribution of each term in 
${A_{20}^{\rm{E} 2}}$, which 
 is estimated by the volume integral of ${A_{20}^{\rm{E} 2}}$ within a given sphere
 enclosed by certain radius.
 It can be seen that
 the contribution of the hydrodynamic and gravity parts (indicated by 
"hyd $\&$ grav") is prominent for radius outside $\sim 1000$ km, which stems from 
exploding regions with large kinetic energy as mentioned above.
 Note that the secular drift observed in the increase-type 
 waveform may come from ambiguities in estimating the gravitational potential in 
 the stress formula as pointed out by \citet{dimmel2002}. 
To exclude such potential issues, we plot the waveform calculated 
 by the first-moment 
of momentum-density formalism (e.g., \citet{finn}). As is shown in Figure \ref{fig15},
 the increasing trend is also seen 
 in the waveform estimated by the first-moment formalism (green line).
As a side-remark, a more smoother curve is obtained for the 
 stress formula (red line), about which M\"onchmeyer et al. (1991) pioneeringly 
 mentioned that the numerical evaluation of the time-derivative sometimes 
makes the waveform noisy.\footnote{This may be the reason why the 
 stress formula has been often employed in supernova researches so far.}

\subsubsection{Cancellation type}

\begin{figure}[htbp]
    \centering
    \includegraphics[width=.44\linewidth]{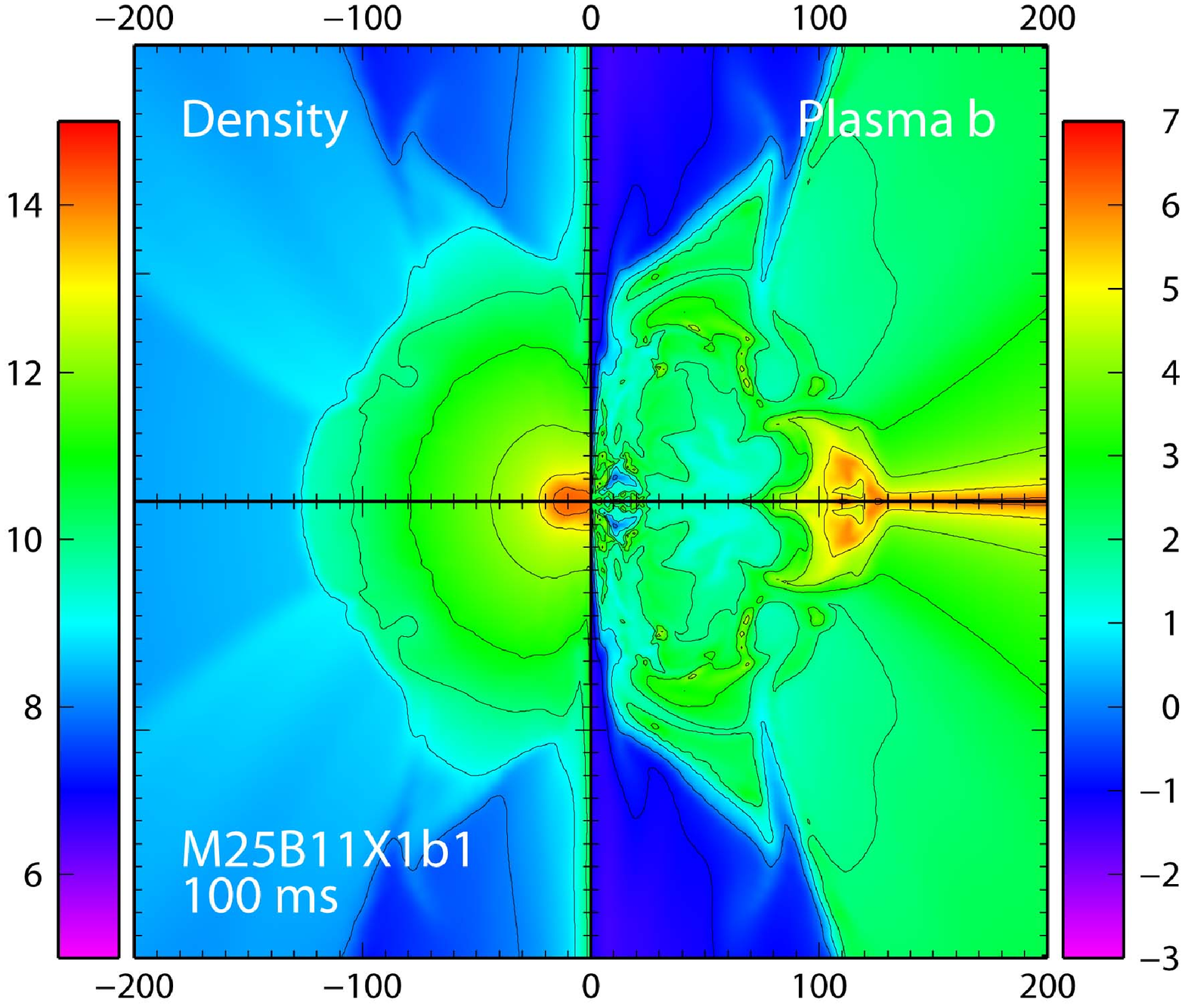}\\
    \includegraphics[width=.44\linewidth]{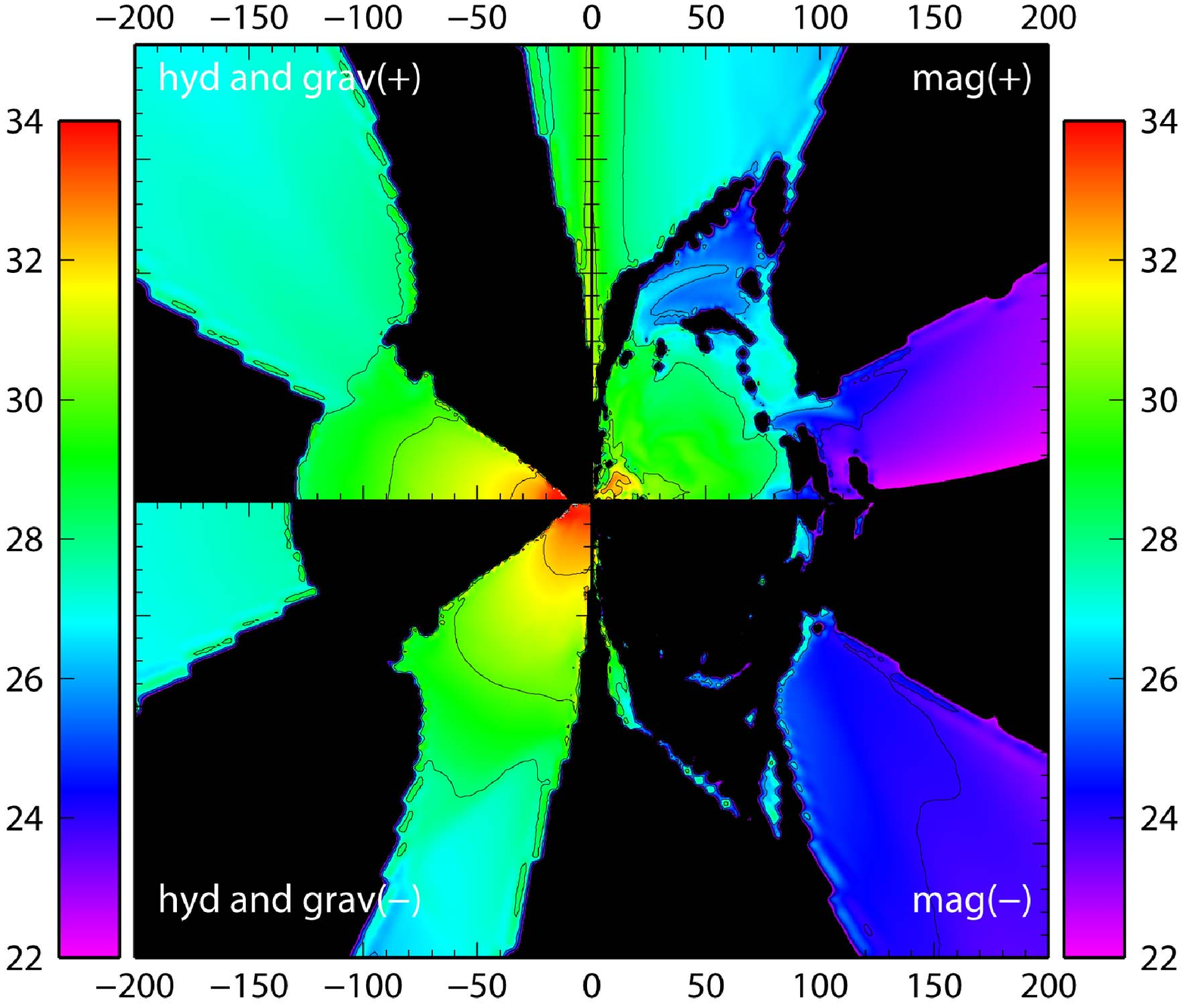}\\
    \includegraphics[width=.44\linewidth]{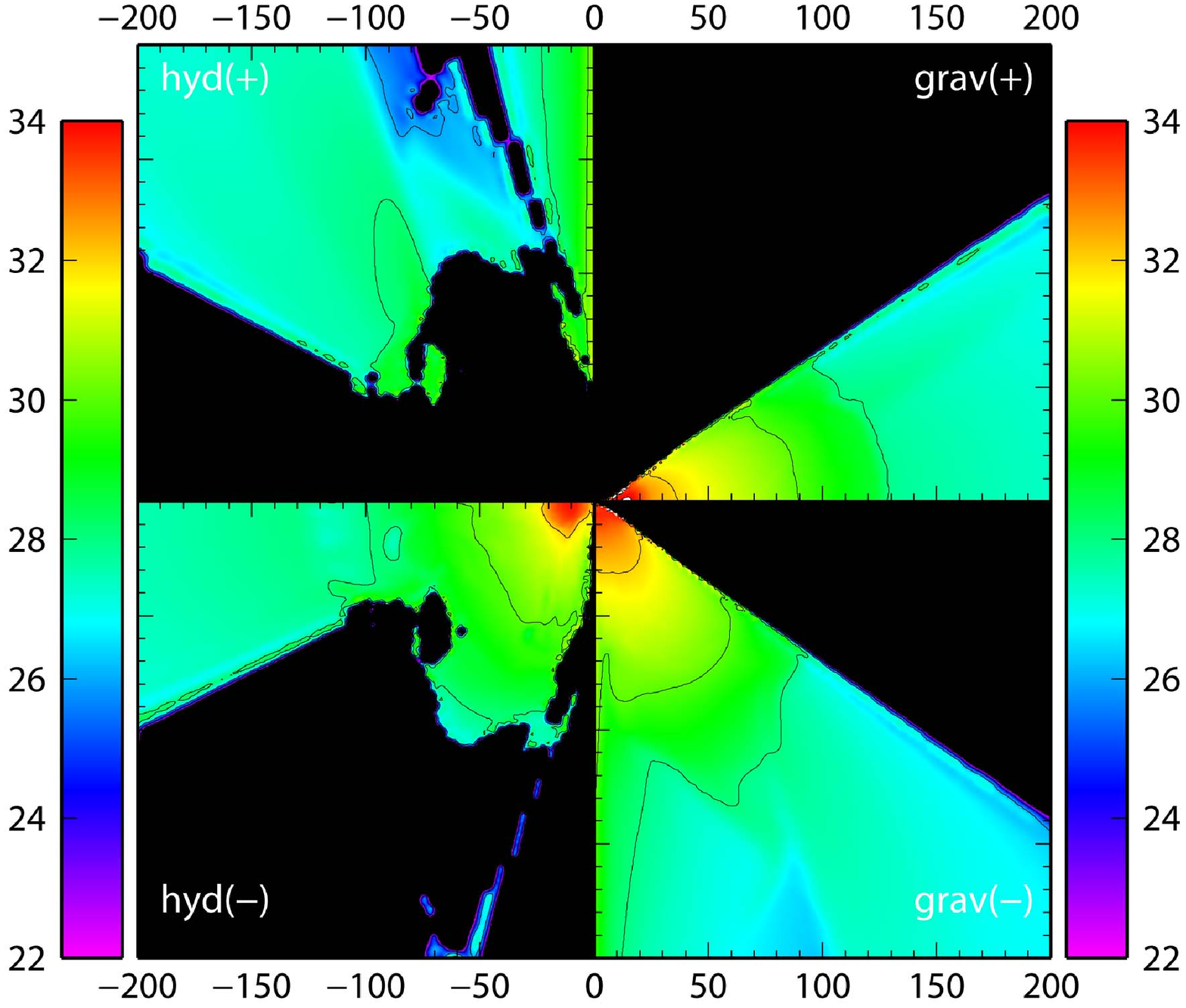}
    \caption{
 The top panel shows distributions of logarithm of density [$\gpcmc$]
 (left-half) and logarithm of $\beta$ 
(right-half) for model B11X1$\beta$1 at 100 ms after bounce. 
 Like Figure \ref{fig:inc_cont}, the middle panel shows the sum of the hydrodynamic 
 \& gravitational part (left-half), while the magnetic contribution is shown
 in the right-half panel.
 The bottom panel is 
 for the hydrodynamic (left-half) and gravity part (right-half), respectively. 
The side length of each plot is 400(km)x400(km). Compared to Figure 2, 
note that a more central region is focused here because the contribution to the 
GWs there are more important for the cancellation-type waveform. }
    \label{fig:cancel_cont}
\end{figure}

Now we proceed to focus on the cancellation-type waveform by 
taking model B11X1$\beta$1 as an example.
Top panel in Figure \ref{fig:cancel_cont} shows density 
 distributions (left-half) with the plasma $\beta$ (right-half). 
  Note that the jet head
 of MHD outflows is located at $\sim 400$ km along the pole which sticks out of the
 plots in Figure \ref{fig:cancel_cont} (compare the difference in scales of 
Figure \ref{fig:inc_cont}).
 In fact, the bluish regions (low $\beta$) around the pole have a lower density
 because material there has been already blown up due to the passage of 
MHD-driven shocks. The middle panel of Figure 5 shows the term-by-term
 contribution to ${A_{20}^{\rm{E} 2}}$ as in the right panel of Figure 
\ref{fig:inc_cont}. From the right-half panel, it is shown
 that the magnetic contribution dominantly makes a positive contribution
(labeled by mag(+)), which is also the case of the increasing type as mentioned
 in the previous section.
 Note that the negative contribution from magnetic fields 
(bluish region in the bottom-right-half of the middle panel) 
comes from the regions, where are off-axis from the propagation of the MHD jets.
 In these regions,
 the poloidal components of magnetic fields 
are stronger than the toroidal ones, which makes the 
 negative contribution mainly 
through $- [{b_r}^2 ( 3 \mu^2 -1) + {b_{\theta}}^2 ( 2 - 3 \mu^2)$] 
in equation (\ref{mag}).
 
\begin{figure}[htbp]
    \centering
    \includegraphics[width=.6\linewidth]{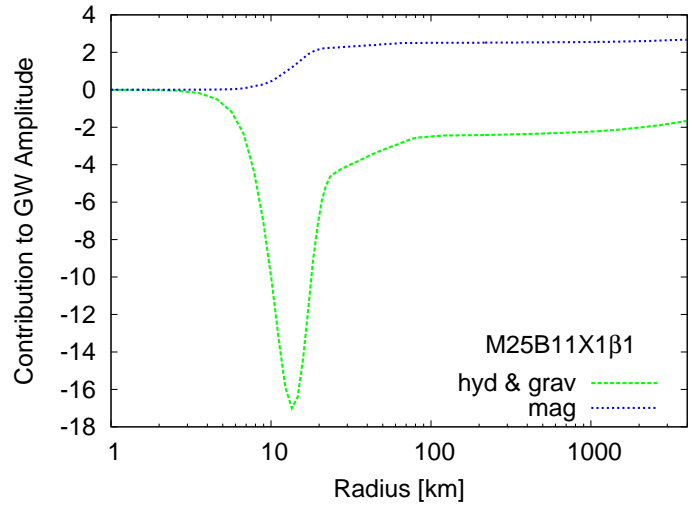}
    \caption{Same as Figure \ref{fig:inc_rad} but for model B11X1$\beta$1.}
    \label{fig:cancel_rad}
\end{figure}

Looking at the sum of the hydrodynamic and gravitational part (left-half
 in the middle panel), a large negative contribution
 comes from regions near in the rotational axis 
(colored by red, bottom-left-half). 
 The bottom panel of Figure \ref{fig:cancel_cont} further shows
 a contribution from the hydrodynamic (left-half) and gravitational part (right-half),
 separately.
 Regarding the hydrodynamic part, the negative contribution is highest 
in the vicinity of the equatorial plane which closely coincides with the oblately 
deformed protoneutron star (colored by red, the bottom-left-half in the bottom panel).
This is because the negative contribution comes from the centrifugal forces 
(e.g., the term related to the rotational energy, 
$ - \rho {v_{\phi}}^{2}$ in equation (\ref{quad})). For the gravity part (right-half
 in the bottom panel), a big negative contribution comes from regions in the vicinity of 
 the rotational axis (bottom-right-half). This comes from the term of 
$- r \partial_{r} \Phi (3 \mu^2 -1)$ in equation (\ref{quad}), which is the radial 
 gradient of the gravitational potential. Remembering that
 $\partial_{r} \Phi > 0$ is generally satisfied in self-gravitating objects, 
the sign of this term is determined by $\mu = \cos \theta$ which is a directional 
cosine measured from the rotational axis. As a result, the gravity part makes 
 a negatively contribution in the vicinity of the rotational axis ($\mu \sim 1$).
  These two factors coming both from the hydrodynamic and gravity part
make the negative contribution to the GW amplitudes. In fact,
 the negative contribution is highest near at the surface of 
the protoneutron star ($\sim 10$ km in Figure \ref{fig:cancel_rad}).
 Outside it, the magnetic part 
 becomes almost comparable to the sum of the hydrodynamic and gravity part,
 which makes the cancellation type shown in the right panels of Figure 
\ref{fig:hakei}.

\clearpage
\subsection{Spectrum Analysis}
\label{C}
\begin{figure}[htbp]
    \centering
    \includegraphics[width=.49\linewidth]{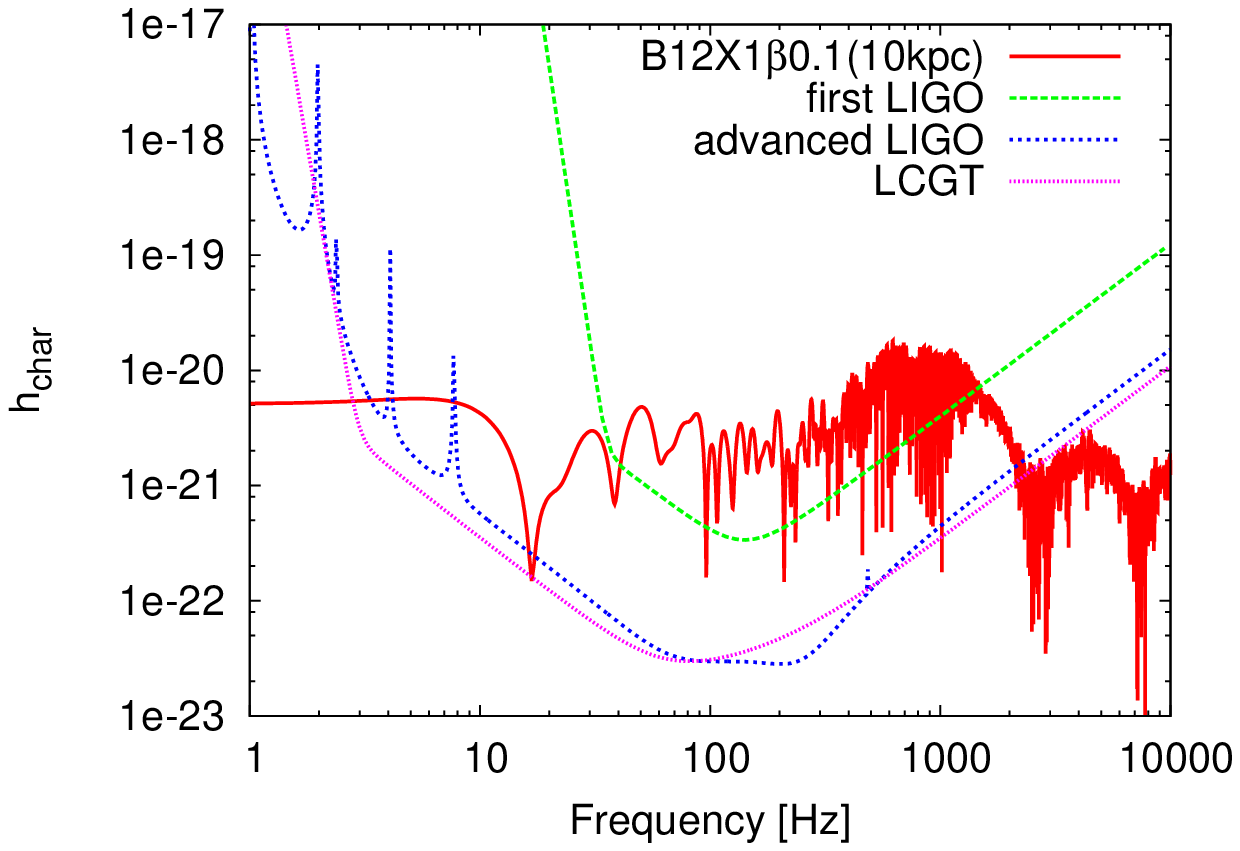}
    \includegraphics[width=.49\linewidth]{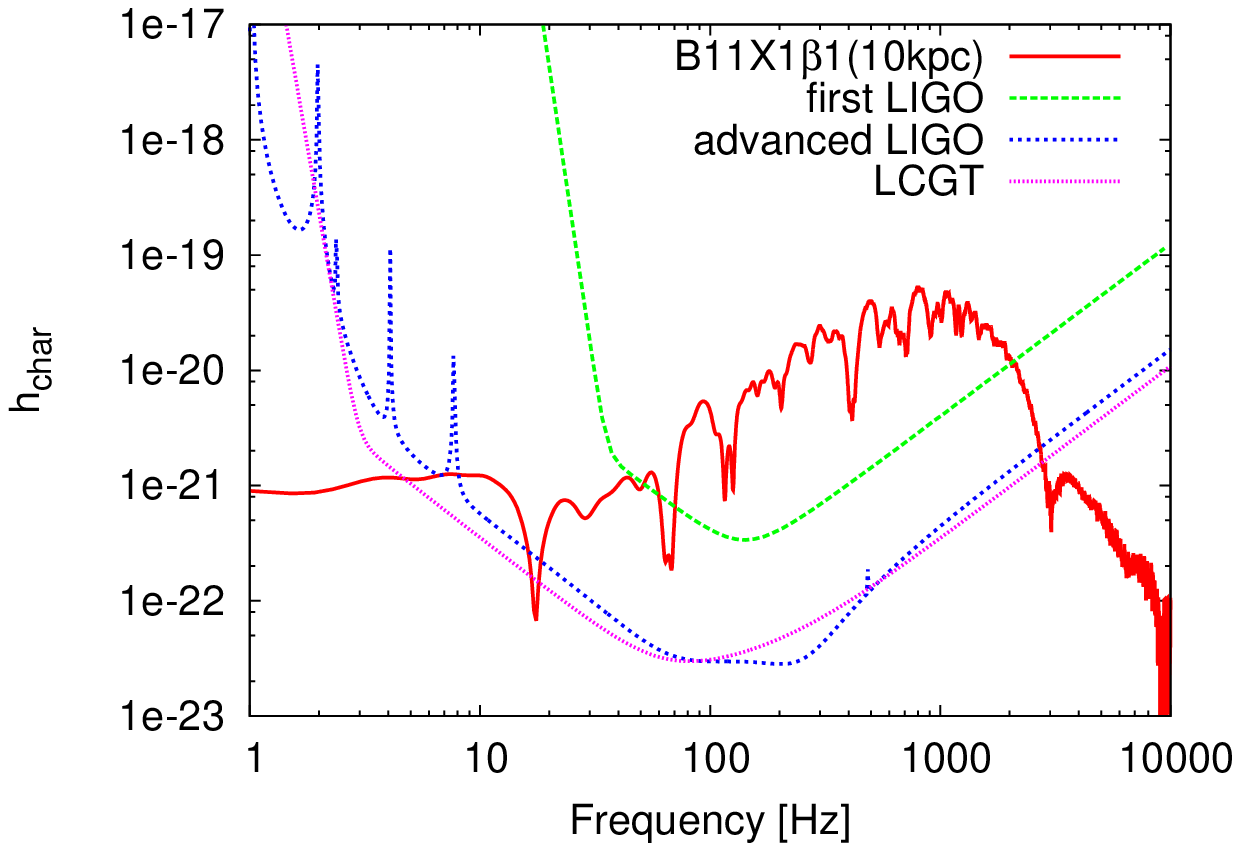}\\
    \includegraphics[width=.49\linewidth]{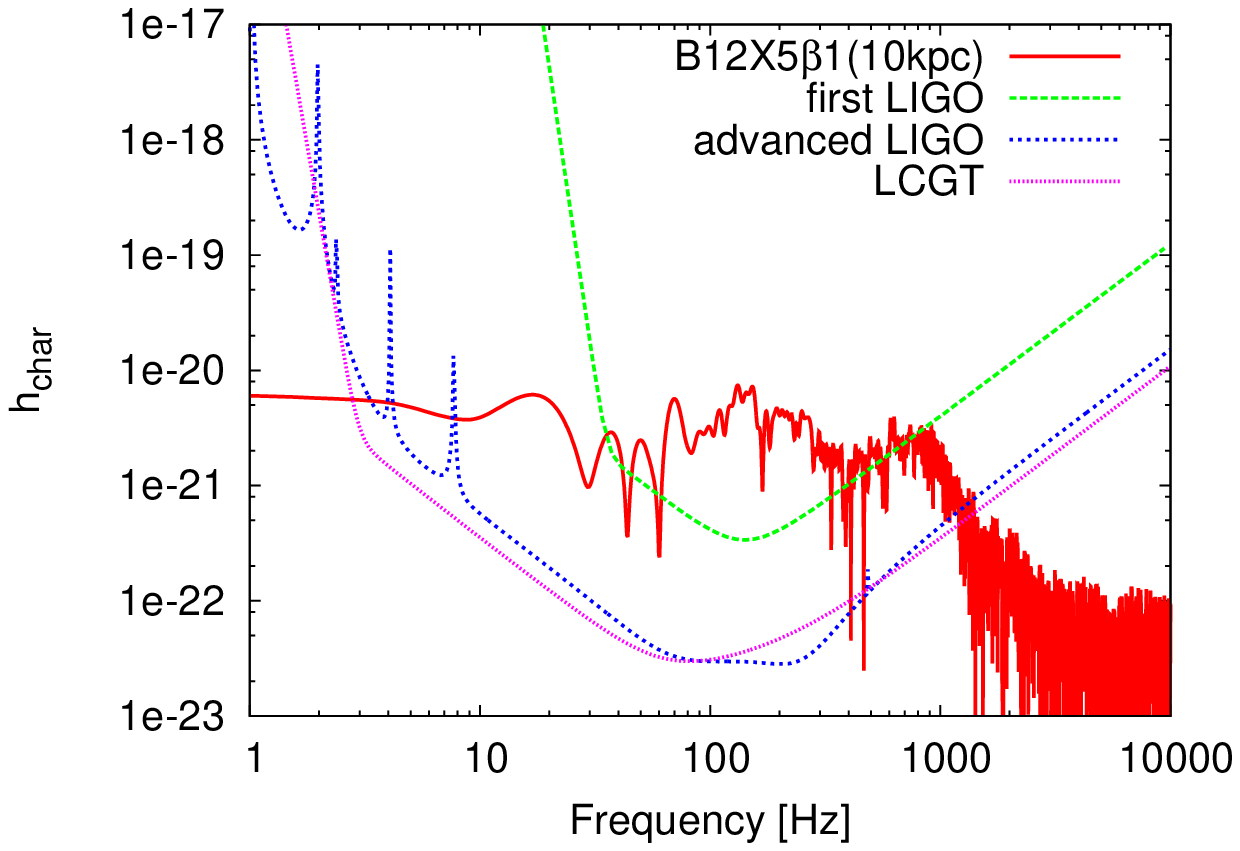}
    \includegraphics[width=.49\linewidth]{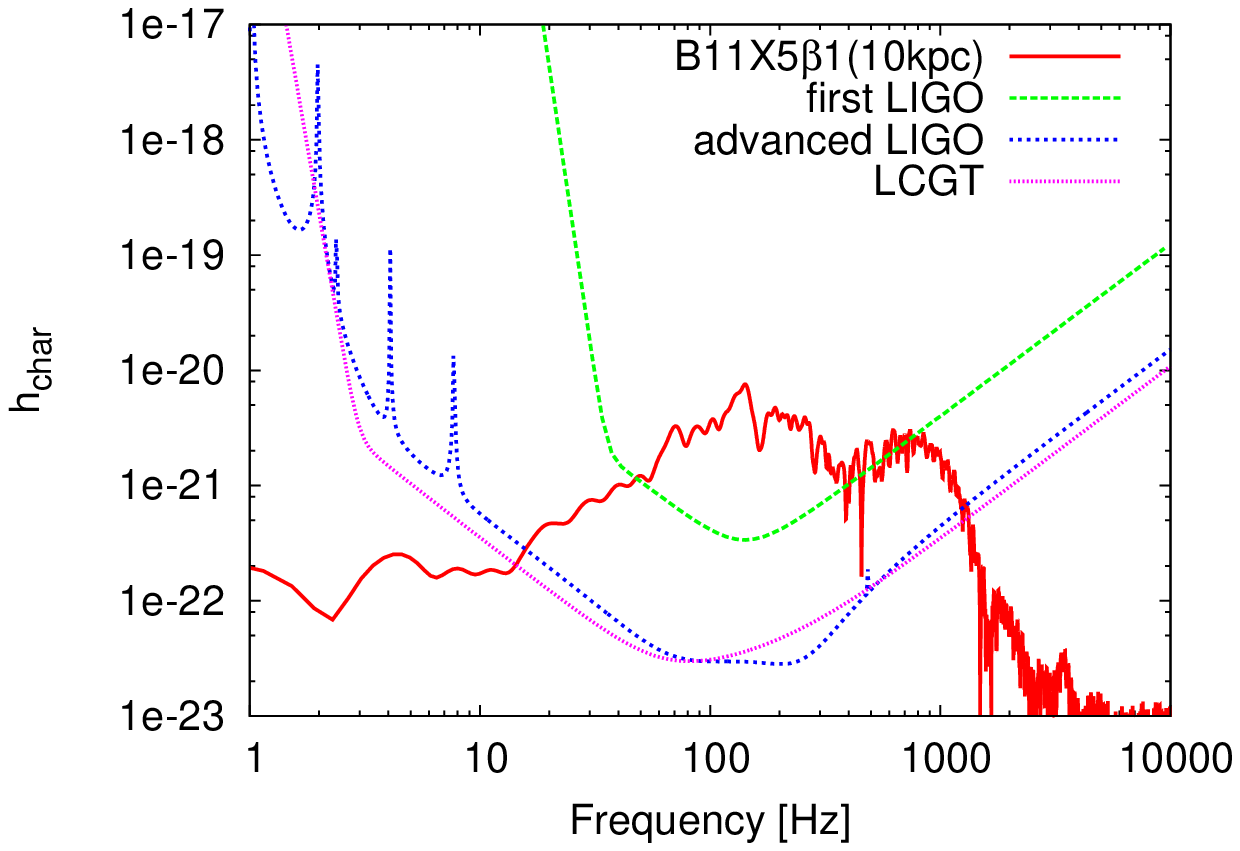}\\
    \caption{
 Gravitational-wave spectrum for representative models with the 
 expected detection limits of the first LIGO \citep{firstligonew}, 
the advanced LIGO \citep{advancedligo}, Large-scale
 Cryogenic Gravitational wave Telescope (LCGT) \citep{lcgt}, and Fabry-Perot type 
 DECIGO \citep{fpdecigo,kudoh}.
It is noted that $h_{\rm char}$ is 
the characteristic gravitational wave strain defined in \citet{flanagan}. 
The supernova is assumed to be located at the distance of 10 kpc.}
    \label{fig:spectrum}
\end{figure}

Now we move on to perform a spectral analysis (Figure 
 \ref{fig:spectrum}). Both for the two types 
(left panel (increasing-type), right panel (cancellation-type)),
 the peak amplitudes in the spectra are around $1$ kHz, which comes 
 from the GWs near at bounce. It can be seen that the spectra for 
lower frequency domains (below $\sim 100$ Hz) are much 
 larger for the increasing type (left panels) compared to the cancellation type 
(right panels). This reflects a slower temporal variation of the secular drift 
inherent to the increase-type waveforms (e.g., Figure \ref{fig:hakei}).
 
As a measure to characterize the dominance in the lower frequency domains, 
we define $\tilde{h}_{\rm low}$, which represents average amplitudes below $100$ Hz
 (see Table \ref{fig:hpeak}). Although the peak amplitudes, $\tilde{h}_{\rm peak}$,
 in the spectra have no clear correlation with the two types, we point 
 out that the final GW amplitudes (the first column) and 
 the  $\tilde{h}_{\rm low}$ (the third column)
  are much larger for the increasing type (colored by yellow) compared to the cancellation type (colored by light blue). 
 In Figure 7, the peak amplitudes near $1$ kHz are, irrespective of the two types,
 marginally within the detection 
limits of the currently running detector of the first LIGO and the detection 
seems more feasible for the next-generation detectors  
such as LCGT and the advanced LIGO for a Galactic supernova. 
 It is true that the GWs in the low frequency domains mentioned above are
 relatively difficult to detect due to seismic noises, but
 a recently proposed future space interferometers like 
Fabry-Perot type DECIGO is designed to be sensitive in the frequency regimes
\citep{fpdecigo,kudoh}.
The sensitivity curve of the detector is plotted with the black line in Figure \ref{fig:spectrum}. 
 Our results suggest that 
these low-frequency signals, if observed, could be one 
 important messenger of the increase-type waveforms that are likely to be 
associated with MHD explosions exceeding $10^{51}$ erg.

\begin{table}[htbp]
    \centering
    \includegraphics[width=.49\linewidth]{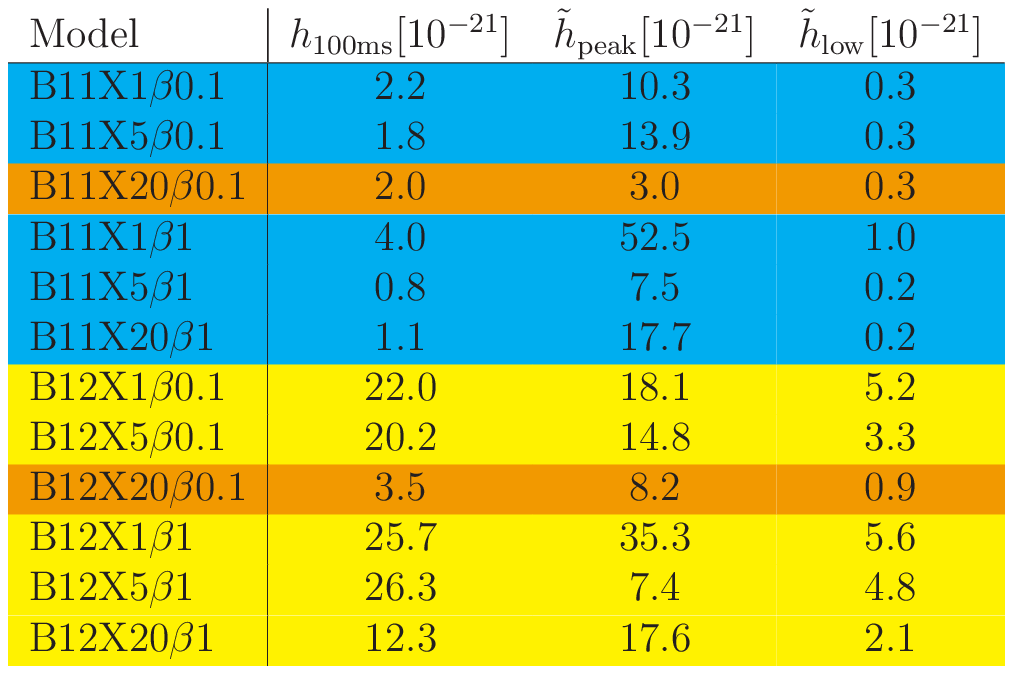}
    \caption{Summary of GW amplitudes for all the models. The colors used for 
 clarify are the same as Table \ref{table1}.
The first column 
 represents the GW amplitudes when we terminated the 
 simulation (100 ms after bounce).
 The second column, $\tilde{h}_{\rm peak}$, is the peak GW amplitudes in the spectra.
The third column,  $\tilde{h}_{\rm low}$, is the average amplitudes
 below 100 Hz. The supernova is assumed 
to be located at the distance of 10 kpc.}
    \label{fig:hpeak}
\end{table}

\section{Summary and Discussion \label{sec5}}

By performing a series of two-dimensional SRMHD simulations, we studied signatures 
 of GWs in the MHD-driven core-collapse supernovae.
In order to extract the gravitational waveforms, we presented
 a stress formula including contributions both from magnetic fields
 and special relativistic corrections.
By changing the precollapse magnetic fields and initial angular 
 momentum distributions parametrically, we computed twelve models.  As for the microphysics, 
a realistic equation of state was employed and the neutrino cooling was 
taken into account via a multiflavor neutrino leakage scheme.
 With these computations, we found that the total GW amplitudes 
show a monotonic increase after bounce for models with a strong precollapse 
magnetic field ($10^{12}$G) also with a rapid rotation imposed.
 We pointed out that this trend stems both 
from the kinetic contribution of MHD outflows with large radial velocities and
 also from the magnetic contribution dominated by the toroidal magnetic fields 
 that predominantly trigger MHD explosions. 
For models with weaker initial magnetic fields, the total GW amplitudes 
after bounce stay almost zero, because the contribution 
from the magnetic fields cancels with the one from the hydrodynamic counterpart.
 These features can be clearly understood with a careful analysis on the explosion 
dynamics. It was pointed out that the GW signals with an increasing trend, possibly visible 
to the next-generation 
detectors for a Galactic supernova, 
 would be associated with MHD explosions exceeding $10^{51}$ erg.

\begin{figure}
  \begin{center}
    \begin{tabular}{cc}
      \resizebox{55mm}{!}{\includegraphics{fig8left.eps}} &
      \resizebox{55mm}{!}{\includegraphics{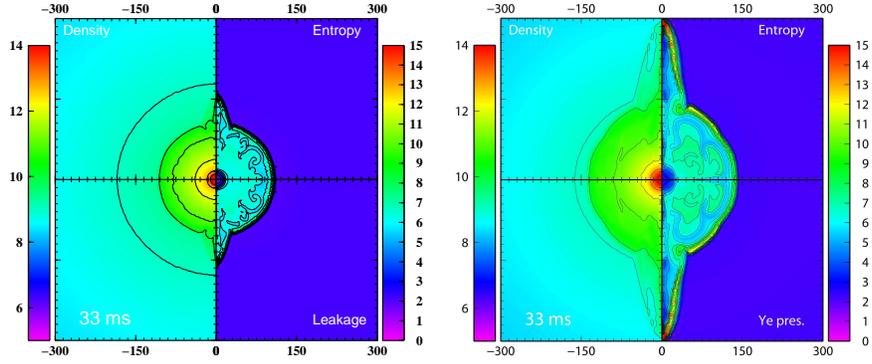}} \\
    \end{tabular}
   \caption{Snapshots at 33 ms after bounce for model B12X20$\beta0.1$ 
in which the leakage scheme (left panel) or the $Y_e$ prescription (right panel) is 
 employed, respectively. In each panel, density (logarithmic, left-half) 
and entropy (right-half) distributions are shown. 
The side length of each plot is 600x600(km).}
  \label{fig11}
  \end{center}
\end{figure}

\begin{figure}
  \begin{center}
    \begin{tabular}{ccc}
      \resizebox{50mm}{!}{\includegraphics{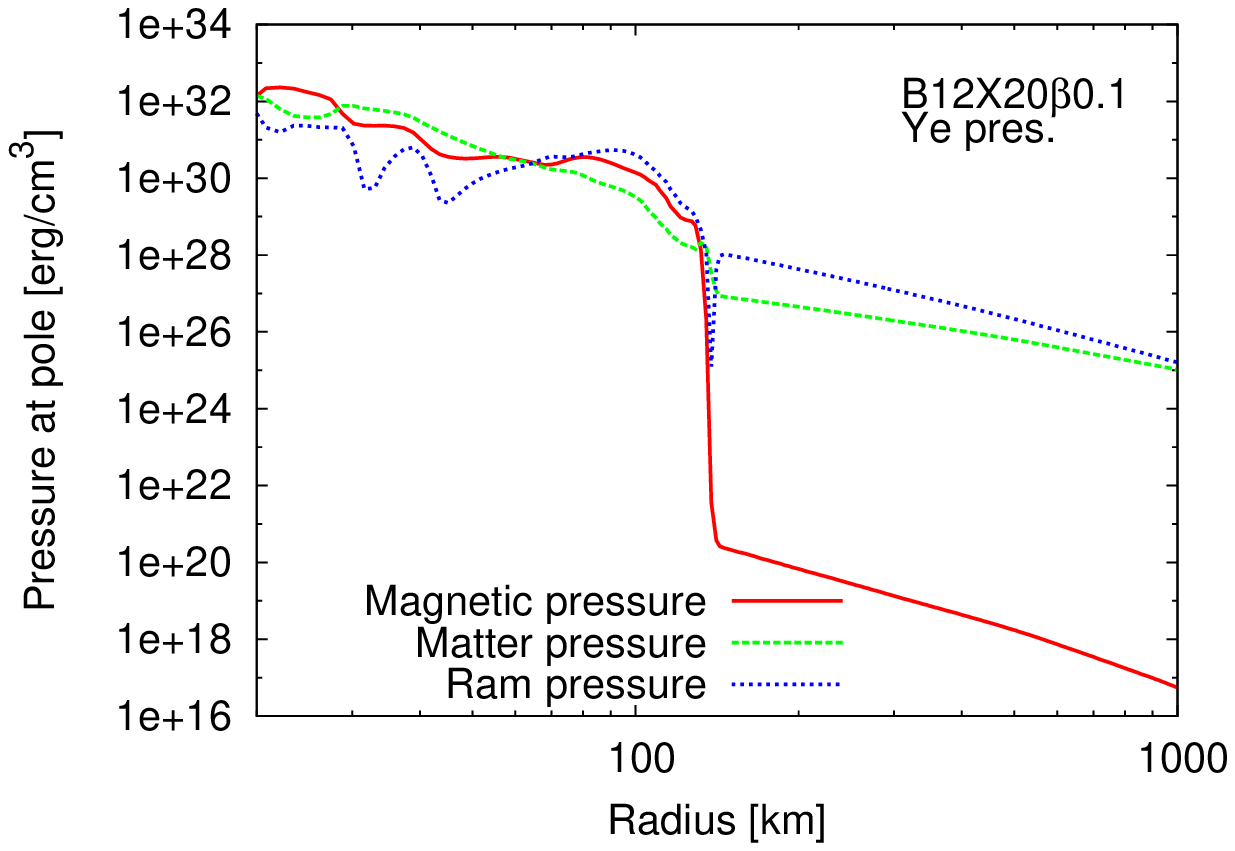}} &
      \resizebox{50mm}{!}{\includegraphics{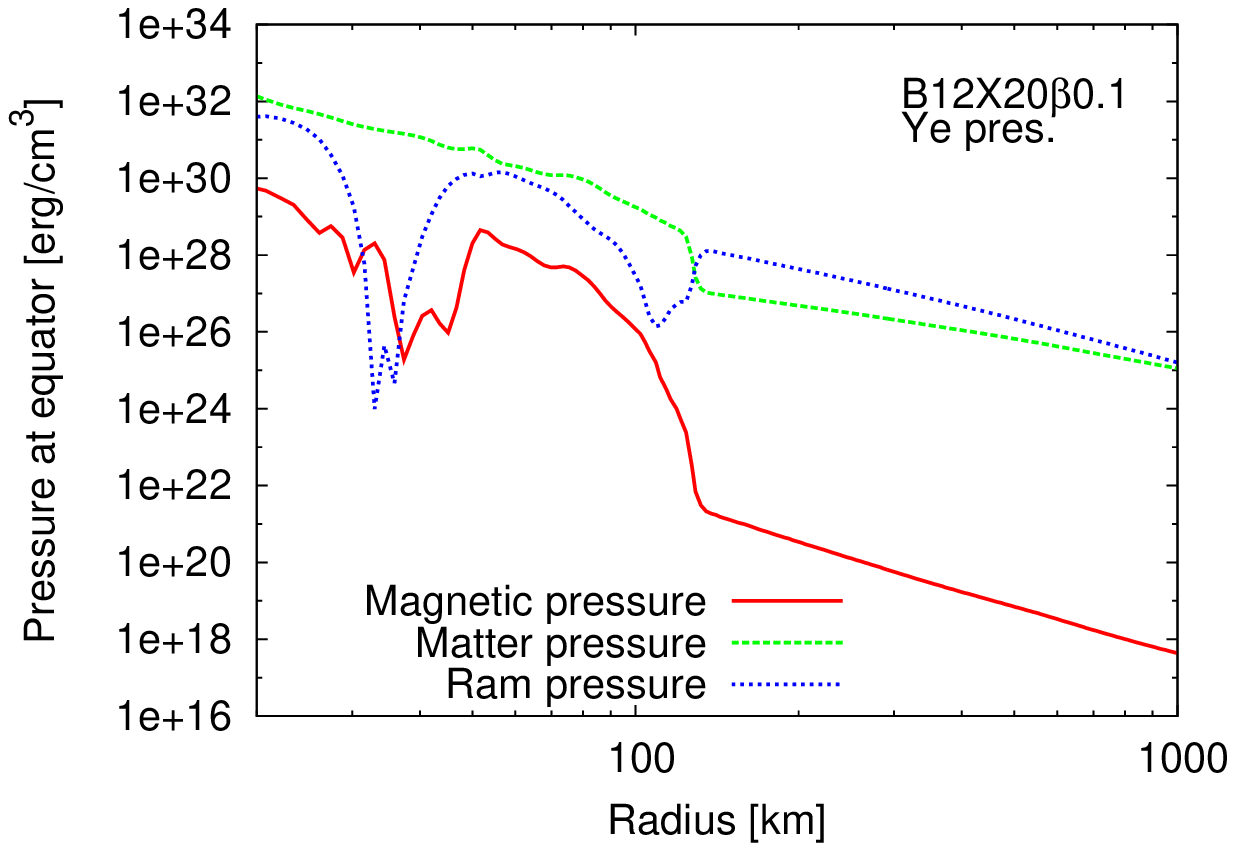}} &
      \resizebox{50mm}{!}{\includegraphics{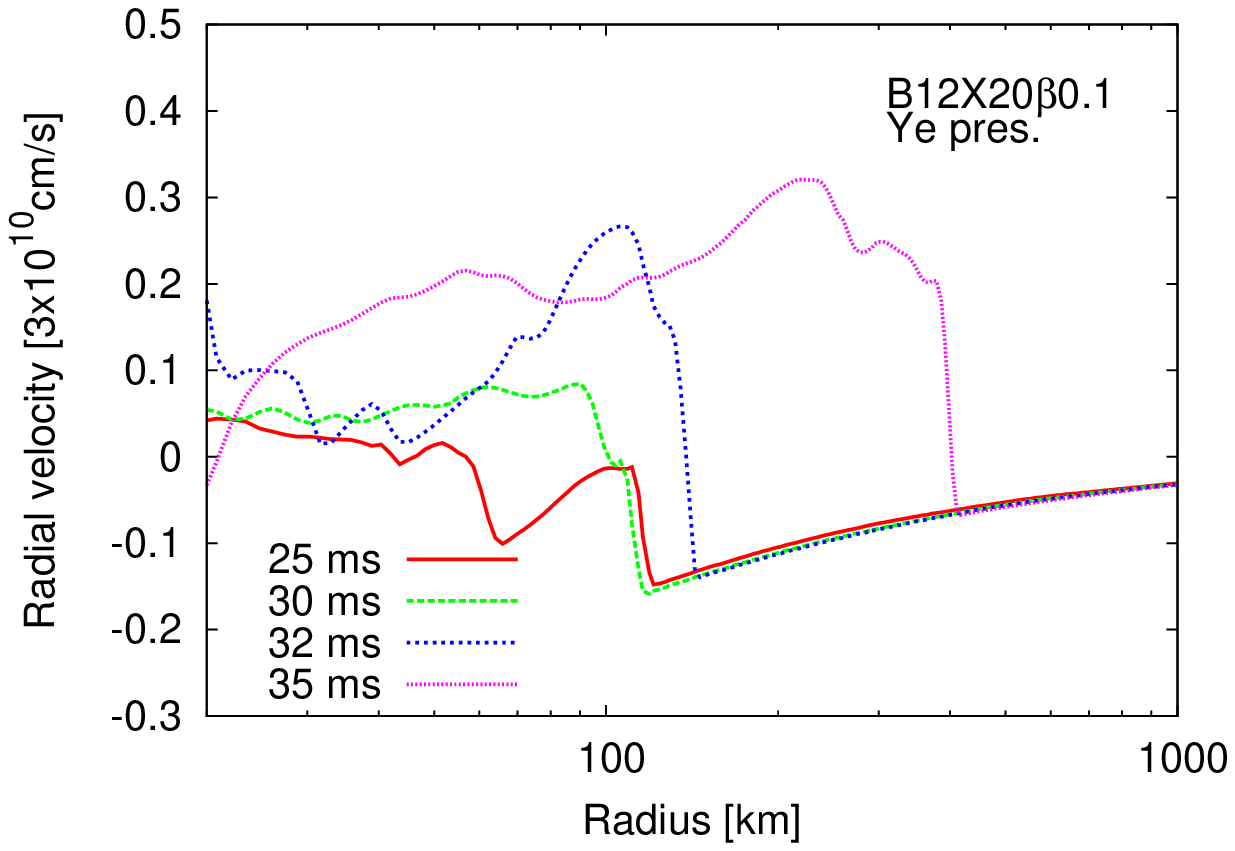}}\\
      \resizebox{50mm}{!}{\includegraphics{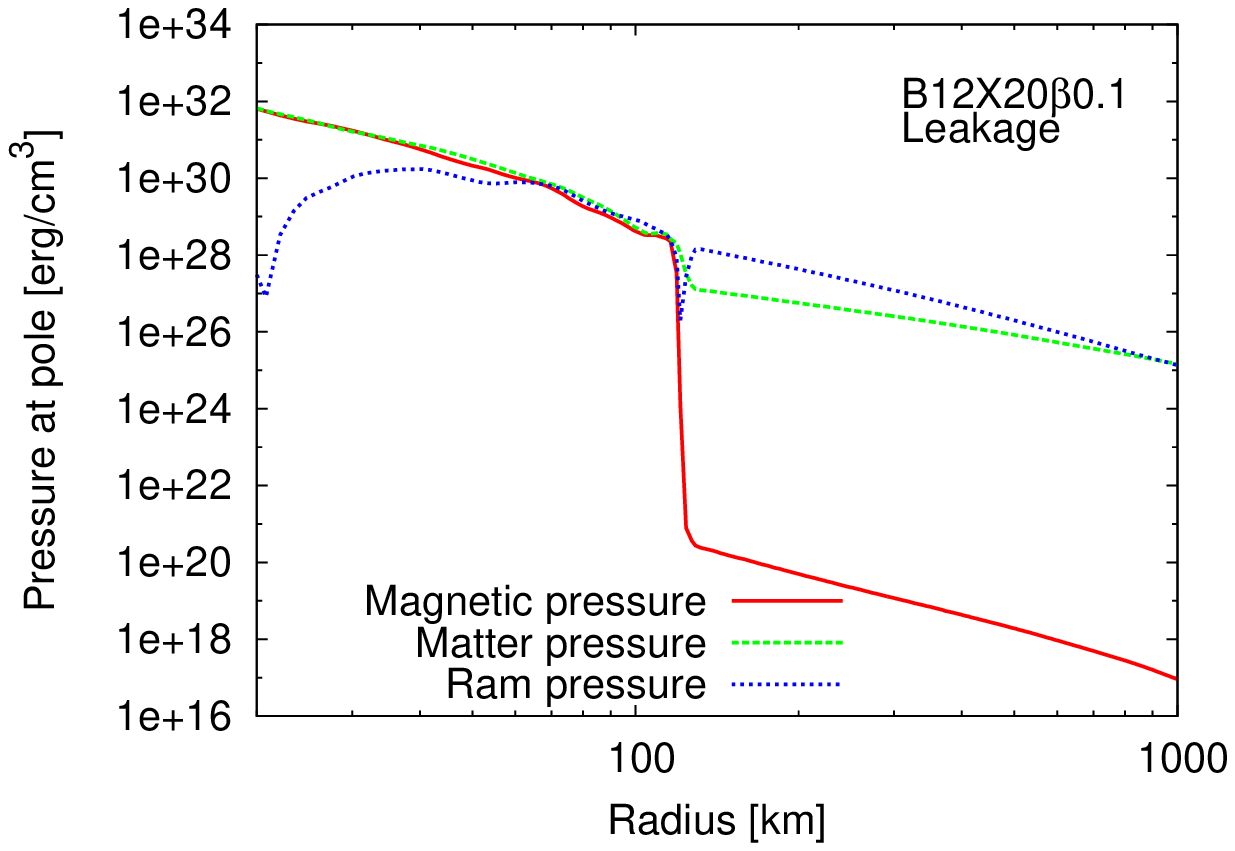}} &
      \resizebox{50mm}{!}{\includegraphics{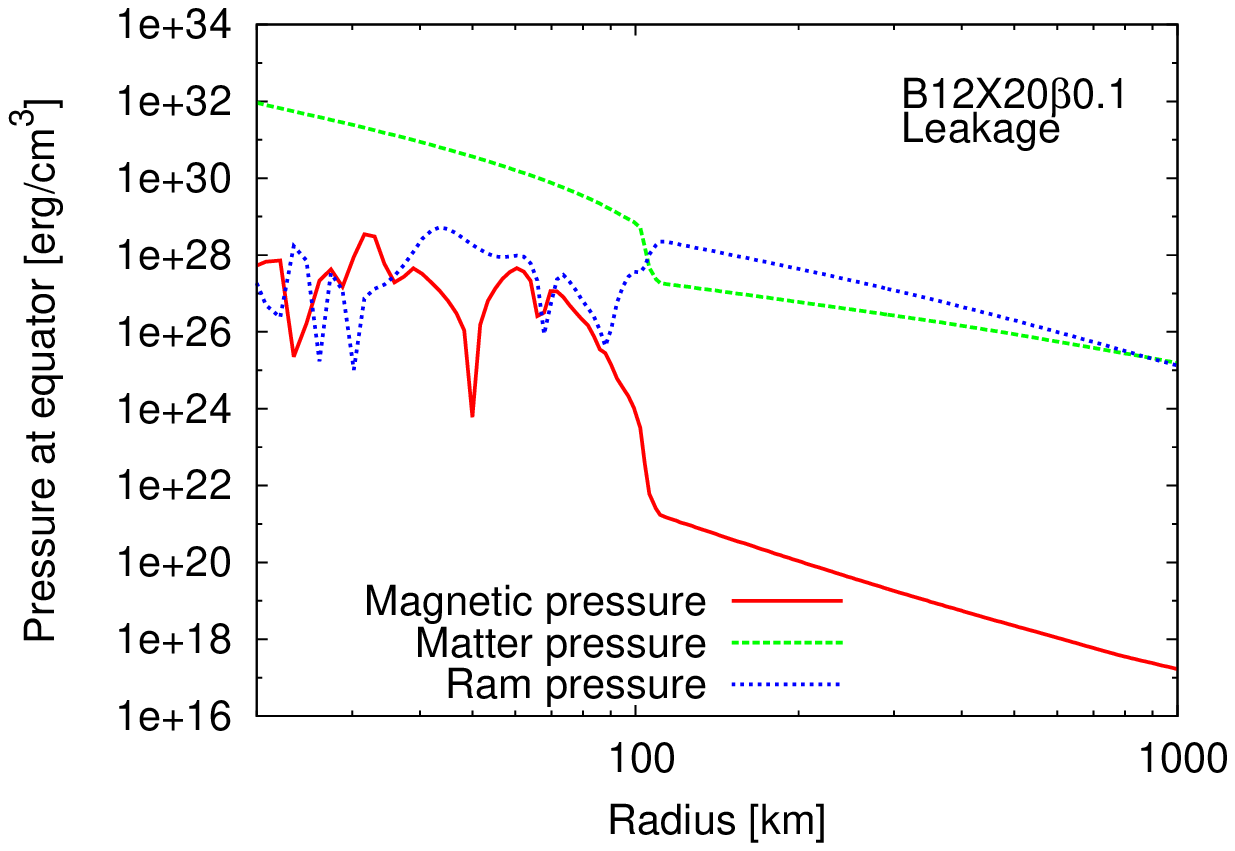}} &
      \resizebox{50mm}{!}{\includegraphics{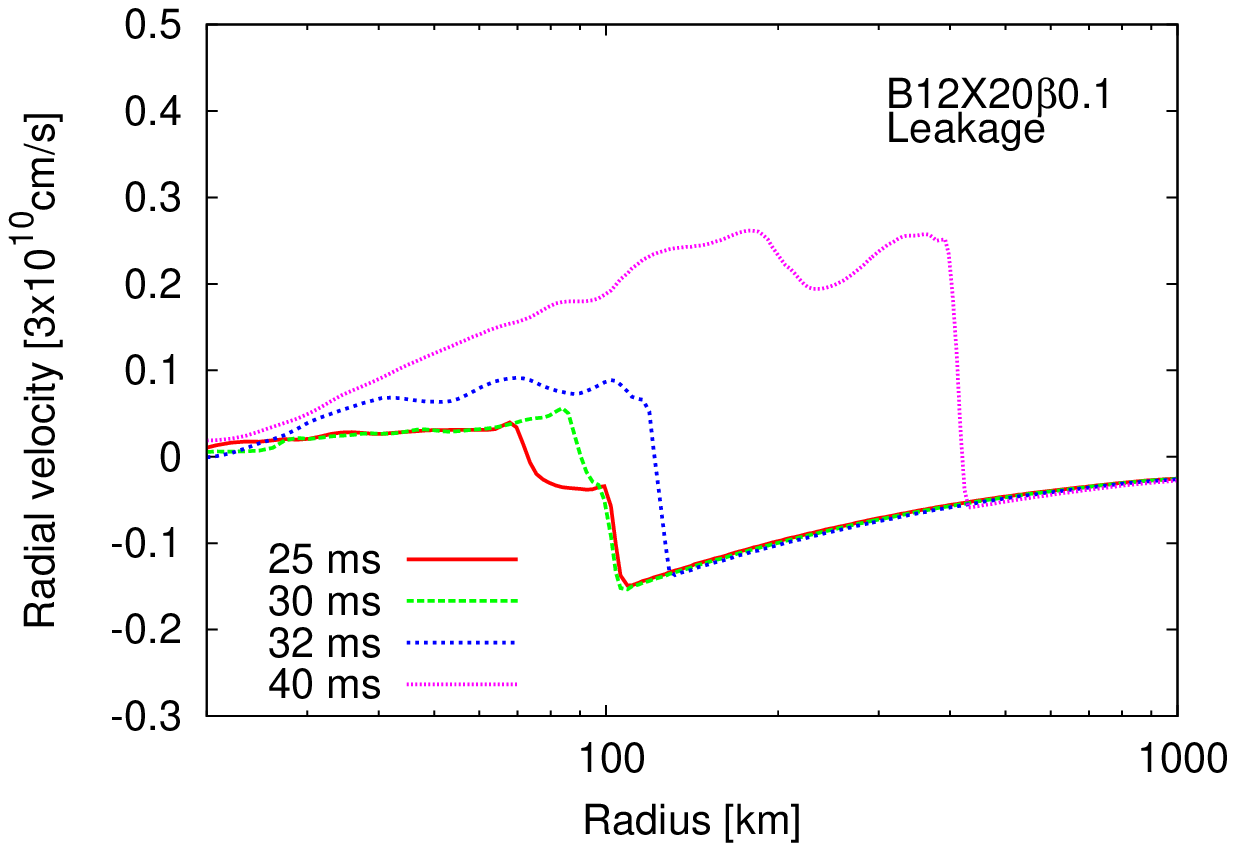}}\\
    \end{tabular}
   \caption{Left and middle panels show magnetic pressure (red line) vs. ram 
pressure (blue line) for model 
B12X20$\beta 0.1$ around 32 ms after bounce with the $Y_e$ prescription (top panels) or the leakage scheme (bottom
 panels) along the polar axis (left panel)
 or the equatorial plane (middle panel). Matter pressure is 
shown by green line as a reference. The right panels show
  velocity profiles along 
 the pole near after the stall of the bounce shock (red lines). Both in the two
 different deleptonization schemes,
 the MHD-driven explosions are indeed obtained.}
  \label{fig12}
  \end{center}
\end{figure}

\begin{figure}[htb]
  \begin{center}
    \begin{tabular}{c}
      \resizebox{66mm}{!}{\includegraphics{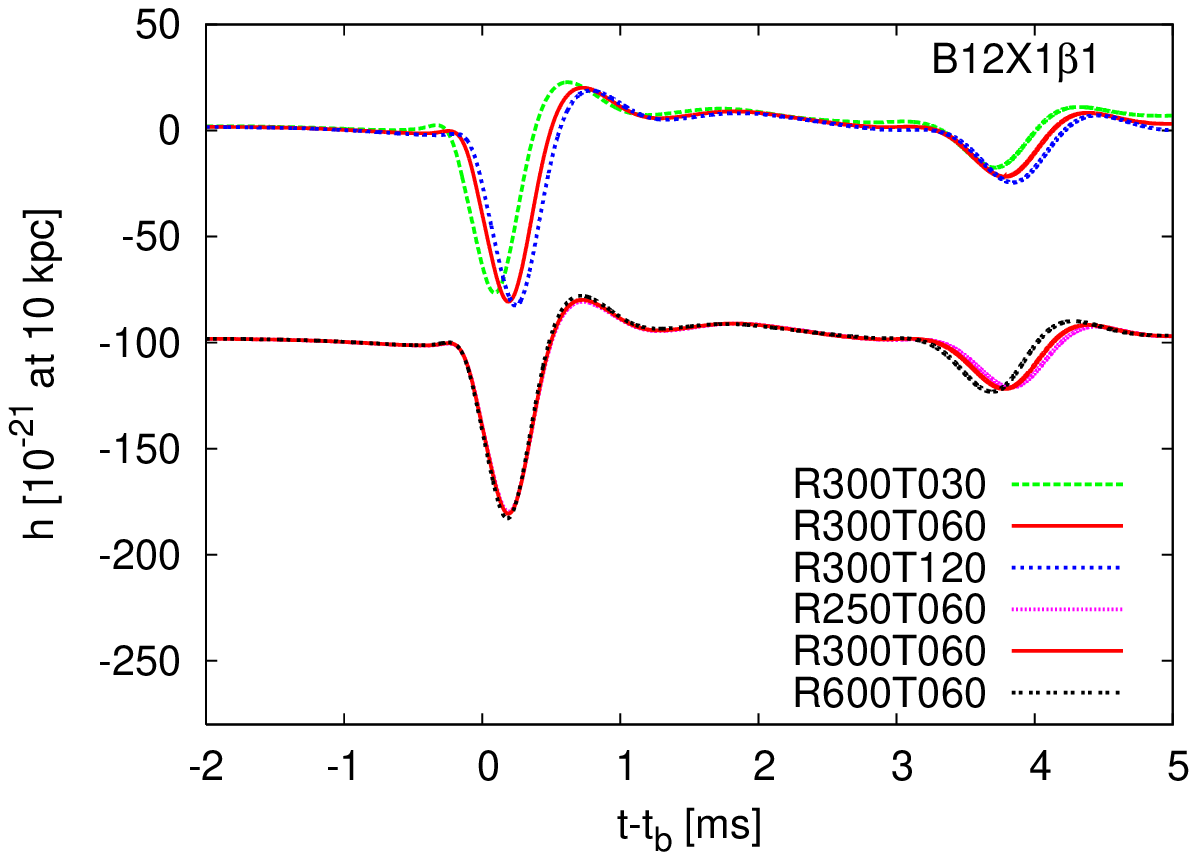}}
    \end{tabular}
   \caption{Gravitational waveforms for model B12X1$\beta1$ near 
bounce with different grid points. Three lines starting from 0 
(at $t - t_{\rm b}$ = -2 ms with $t_{\rm b}$ being the epoch of bounce),
 correspond to models with different angular grid points 
(30 (green), 60(red), 120(blue)) while fixing the radial grid points to be 300.
 The bottom three lines are set to start from -100 in the GW amplitudes (just for 
 convenience), and they correspond to models with 
different radial grid points (250 (pink), 300 (orange), 600 (brown)) 
while fixing the lateral grid points to be 60. 
Note that the fiducial set employed in this work is 300(r)x60($\theta$).}
  \label{fig14}
  \end{center}
\end{figure}


Although the presented simulations have utilized the leakage scheme to approximate 
  the deleptonization, it would be more accurate (especially before bounce)
 to employ a formula developed by \citet{matthias05},
 which was designed to fit 1D Boltzmann results. 
Figure \ref{fig11} shows snapshots at around 33 ms after 
bounce for model B12X20$\beta0.1$ in which the leakage scheme 
(left panel) or the $Y_e$ prescription (right panel) is employed, respectively.
Note that ``G15'' is taken in our simulation until bounce among the parameter sets in \citet{matthias05}\footnote{The inner-core mass at bounce for the employed 
 parameter set is 0.6 $M_{\odot}$ for our non-rotating 25 $M_{\odot}$ progenitor. 
This value is higher than that obtained in GR simulations (0.45-0.55 $M_{\odot}$) 
in \citet{dimm08}. This may be because the pseudo GR potential 
employed in this work underestimates the GR gravity, which could potentially
 lead to a large inner-core mass.}
and is switched to the leakage scheme at the postbounce phase.
 As is shown, the shock revival also occurs for the 
model with the $Y_e$ prescription (right panel). Figure \ref{fig12} 
depicts the magnetic pressure (red line) vs. ram pressure (blue line) 
along the polar axis (left panel) or the equatorial plane (middle) near 
the rebirth of the stalled shock for the model with the $Y_e$ prescription
 (top panels) or the leakage scheme (bottom panels).
 For the equator, the magnetic pressure is much less than the ram pressure
 (middle panel), while the magnetic pressure amplified by the field wrapping 
 along the pole becomes as high as the ram pressure of the 
infalling material at the shock front, leading to the MHD-driven shock formation 
(see right panels). Regardless of the two different deleptoniaztion schemes,
 these important features associated with the MHD explosions are shown to 
 be quite similar.

 Now we mention a comparison between the obtained results and relevant 
MHD simulations. Model R4E1CF in \citet{scheid} whose precollapse rotational parameter 
is $\beta = 0.5$ \% with a uniform rotation imposed, and whose initial poloidal 
magnetic field is set to $10^{12}$ G, is close to our model B12X20$\beta0.1$. 
 From their Figure 23,
 the jet propagates to $\sim$ 300 km along the rotational axis
 at around $\sim 18$ ms after bounce.
 In our counterpart model, 
the MHD-driven shock revives after around 30 ms after bounce, and 
 it reaches to $300$ km at around $\sim$ 10 ms, which is equivalent to $\sim$ 
40 ms after bounce.
  Considering that our model ($\beta = 0.1$\%) is a slower rotator than model
 R4E1CF ($\beta = 0.5$\%), the delay of the shock revival for our model seems reasonable.
 Model s20A1B5-D3M12 in \citet{pablo} whose precollapse angular velocity
 is 4 rad/s (the rotational parameter should be close to $\beta= 0.1$ \%)
 with uniform rotation and whose initial poloidal magnetic field 
is $10^{12}$ G, is close to our model B12X20$\beta0.1$. From 
their Figure 10, the MHD jet propagates to $\sim$ 240 km at 51 ms after bounce.
The employed EOS is the same as ours (the Shen EOS), while the deleptonization
 scheme taken in their study was the $Y_e$ formula \citep{matthias05}. 
As mentioned above, 
 the dynamics is rather close to our corresponding model. 
Among the computed models in \citet{burr07}, model M15B12DP2A1H which 
 has a precollapse angular velocity of $\pi$ rad/s 
(the rotational parameter should be close to $\beta = 0.06$\%) 
with initial dipolar magnetic field of $10^{12}$ G, is close 
to our model B12X20$\beta0.1$. The interval before the launch of the MHD shock for 
their model is 80 ms after bounce 
 (e.g., their Table 1) is much later than our model (28 ms after bounce).
 This may be due to the larger initial angular momentum ($\beta = 0.1 \%$) 
assumed in our study.

 Model A3B3G5-D3M13 in Obergaulinger et al. (2005) which 
 has a rotational parameter of $\beta=0.9$\% with a differential rotation imposed
 (the radial cut off is 500 km) and the initial poloidal magnetic field is 
 $10^{13}$ G, is closer to our model B12X1$\beta0.1$.
 The MHD jet reaches to 
 500 km at around 7 ms after bounce, which is also the case of our counterpart 
model. As discussed above, our results are 
compatible to the ones obtained in the relevant foregoing results.

The major limitation of this study is the assumption of axisymmetry.
Recently it was reported in three-dimensional (3D)
 MHD core-collapse simulations \citep{simon,scheid}
that the fast growth of the spiral SASI hinders
 the efficient amplification of the toroidal fields, which could suppress 
 the formation of jets rather easily realized in 2D simulations. As a sequel of this study, 
 we plan to investigate the 3D effects in SRMHD.
 Regarding a resolution dependence 
of our results, Figure \ref{fig14} indicates that 
 our standard resolution is adequate to follow the evolution of the computed models.
 However, it is not sufficient at all to capture the 
magneto-rotational instability (MRI, e.g., \citet{balb98}). 
 At least $10-100$ times finer mesh points are
 required for resolving the MRI \citep{MRI},
 which may require some adaptive-mesh-refinement treatment, a very important  
 component that remains to be improved.
If the MRI could be resolved, the increasing-type waveform could emerge also for models 
with weakly initial magnetic fields because a more 
 efficient field amplification could be captured.
 Although the general relativistic effects were treated only by a very
 approximative way, we think that the general relativity should not drastically 
change our results qualitatively, because the central protoneutron stars will not 
collapse to a black hole during our simulation time 
as inferred from a simple argument of the compactness of the inner-core.
As for the microphysics, the neutrino heating is not included 
in this study. However the inclusion of the neutrino heating
 may play a minor role in the waveforms, since
 the timescales before the neutrino-driven explosions set in 
(e.g., \cite{marek,bruenn,suwa} and references therein) are much longer than the 
 MHD explosions.

  As one possible extension of this study, we think it 
interesting to study GWs from anisotropic neutrino radiation in the MHD 
models. Extrapolating the outcome obtained in previous studies (e.g., 
\citet{mueller04,kotake_ray}), 
 we anticipate that the total amplitudes become larger when we include the neutrino 
 GWs. This is because the neutrino GWs may make a positive contribution since the 
 neutrino emissions from the oblately deformed protoneutron stars become 
 stronger toward the polar direction \citep{jamu,kota03a,walder,ott_angle}. If this is really the case, 
the total GW amplitudes 
especially for the increasing type should be much larger, possibly making its detection 
more promising. Furthermore, the neutrino signals from MHD explosions, may 
 have a sharp directional dependence through neutrino oscillations, 
reflecting the aspherical propagation of the shock to the stellar envelope \citep{kawa}. 
Taking a correlation analysis between the neutrino and GW signals could help to reveal 
 the hidden nature of the central engines.
 In fact, several observational proposals have been 
  made in this direction recently \citep{van,aso,leonor}. The MHD-driven core-collapse
 supernovae, albeit rather minor among typical type II supernova explosions,
seem to still contain a number of rich research subjects.

\acknowledgements{We are grateful to K. Sato and S. Yamada
 for continuing encouragements. We also thank to our anonymous referee for a number of
 suggestions which helped us a lot to improve our manuscript. 
Numerical computations were in part carried on XT4 and 
general common use computer system at the center for Computational Astrophysics, CfCA, 
the National Astronomical Observatory of Japan.  This study was supported 
in part by the Grants-in-Aid for the Scientific Research 
from the Ministry of Education, Science and Culture of Japan 
(Nos. 19104006, 19540309 and 20740150).
}

\bibliographystyle{apj} 
\bibliography{srmhdgw}

\begin{thebibliography}{86}
\expandafter\ifx\csname natexlab\endcsname\relax\def\natexlab#1{#1}\fi

\bibitem[{{Abbott et al.}(2005)}]{firstligonew}
{Abbott et al.}, M. 2005, \prd, 72, 122004

\bibitem[{{Ando} \& {the TAMA Collaboration}(2005)}]{tamanew}
{Ando}, M. \& {the TAMA Collaboration}. 2005, Classical and Quantum Gravity,
  22, 881

\bibitem[{{Ardeljan} {et~al.}(2000){Ardeljan}, {Bisnovatyi-Kogan}, \&
  {Moiseenko}}]{arde00}
{Ardeljan}, N.~V., {Bisnovatyi-Kogan}, G.~S., \& {Moiseenko}, S.~G. 2000,
  Astron. Astrophys., 355, 1181

\bibitem[{{Aso} {et~al.}(2008){Aso}, {M{\'a}rka}, {Finley}, {Dwyer}, {Kotake},
  \& {M{\'a}rka}}]{aso}
{Aso}, Y., {M{\'a}rka}, Z., {Finley}, C., {Dwyer}, J., {Kotake}, K., \&
  {M{\'a}rka}, S. 2008, Classical and Quantum Gravity, 25, 114039

\bibitem[{{Balbus} \& {Hawley}(1998)}]{balb98}
{Balbus}, S.~A. \& {Hawley}, J.~F. 1998, Reviews of Modern Physics, 70, 1

\bibitem[{{Bionta} {et~al.}(1987){Bionta}, {Blewitt}, {Bratton}, {Caspere}, \&
  {Ciocio}}]{bionta}
{Bionta}, R.~M., {Blewitt}, G., {Bratton}, C.~B., {Caspere}, D., \& {Ciocio},
  A. 1987, Physical Review Letters, 58, 1494

\bibitem[{{Blanchet} {et~al.}(1990){Blanchet}, {Damour}, \&
  {Schaefer}}]{blanchet}
{Blanchet}, L., {Damour}, T., \& {Schaefer}, G. 1990, \mnras, 242, 289

\bibitem[{{Bruenn} {et~al.}(2010){Bruenn}, {Mezzacappa}, {Hix}, {Blondin},
  {Marronetti}, {Messer}, {Dirk}, \& {Yoshida}}]{bruenn}
{Bruenn}, S.~W., {Mezzacappa}, A., {Hix}, W.~R., {Blondin}, J.~M.,
  {Marronetti}, P., {Messer}, O.~E.~B., {Dirk}, C.~J., \& {Yoshida}, S. 2010,
  ArXiv e-prints

\bibitem[{{Buras} {et~al.}(2006){Buras}, {Rampp}, {Janka}, \&
  {Kifonidis}}]{buras06}
{Buras}, R., {Rampp}, M., {Janka}, H.-T., \& {Kifonidis}, K. 2006, \aap, 447,
  1049

\bibitem[{{Burrows} {et~al.}(2007{\natexlab{a}}){Burrows}, {Dessart}, {Livne},
  {Ott}, \& {Murphy}}]{burr07}
{Burrows}, A., {Dessart}, L., {Livne}, E., {Ott}, C.~D., \& {Murphy}, J.
  2007{\natexlab{a}}, Astrophys. J., 664, 416

\bibitem[{{Burrows} \& {Hayes}(1996)}]{burohey}
{Burrows}, A. \& {Hayes}, J. 1996, Physical Review Letters, 76, 352

\bibitem[{{Burrows} {et~al.}(2006){Burrows}, {Livne}, {Dessart}, {Ott}, \&
  {Murphy}}]{burr06}
{Burrows}, A., {Livne}, E., {Dessart}, L., {Ott}, C.~D., \& {Murphy}, J. 2006,
  Astrophys. J., 640, 878

\bibitem[{{Burrows} {et~al.}(2007{\natexlab{b}}){Burrows}, {Livne}, {Dessart},
  {Ott}, \& {Murphy}}]{burrows2}
---. 2007{\natexlab{b}}, Astrophys. J., 655, 416

\bibitem[{{Cerd{\'a}-Dur{\'a}n} {et~al.}(2008){Cerd{\'a}-Dur{\'a}n}, {Font},
  {Ant{\'o}n}, \& {M{\"u}ller}}]{pablo}
{Cerd{\'a}-Dur{\'a}n}, P., {Font}, J.~A., {Ant{\'o}n}, L., \& {M{\"u}ller}, E.
  2008, Astron. Astrophys., 492, 937

\bibitem[{{Cerd{\'a}-Dur{\'a}n} {et~al.}(2007){Cerd{\'a}-Dur{\'a}n}, {Font}, \&
  {Dimmelmeier}}]{cerd07}
{Cerd{\'a}-Dur{\'a}n}, P., {Font}, J.~A., \& {Dimmelmeier}, H. 2007, \aap, 474,
  169

\bibitem[{{De Villiers} {et~al.}(2003){De Villiers}, {Hawley}, \&
  {Krolik}}]{devi03}
{De Villiers}, J.-P., {Hawley}, J.~F., \& {Krolik}, J.~H. 2003, Astrophys. J.,
  599, 1238

\bibitem[{{Dimmelmeier} {et~al.}(2002{\natexlab{a}}){Dimmelmeier}, {Font}, \&
  {M{\" u}ller}}]{dimm02}
{Dimmelmeier}, H., {Font}, J.~A., \& {M{\" u}ller}, E. 2002{\natexlab{a}},
  Astron. Astrophys., 393, 523

\bibitem[{{Dimmelmeier} {et~al.}(2002{\natexlab{b}}){Dimmelmeier}, {Font}, \&
  {M{\"u}ller}}]{dimmel2002}
{Dimmelmeier}, H., {Font}, J.~A., \& {M{\"u}ller}, E. 2002{\natexlab{b}}, \aap,
  393, 523

\bibitem[{{Dimmelmeier} {et~al.}(2007){Dimmelmeier}, {Ott}, {Janka}, {Marek},
  \& {M{\"u}ller}}]{dimmelprl}
{Dimmelmeier}, H., {Ott}, C.~D., {Janka}, H.-T., {Marek}, A., \& {M{\"u}ller},
  E. 2007, Physical Review Letters, 98, 251101

\bibitem[{{Dimmelmeier} {et~al.}(2008){Dimmelmeier}, {Ott}, {Marek}, \&
  {Janka}}]{dimm08}
{Dimmelmeier}, H., {Ott}, C.~D., {Marek}, A., \& {Janka}, H. 2008, \prd, 78,
  064056

\bibitem[{{Epstein} \& {Pethick}(1981)}]{epst81}
{Epstein}, R.~I. \& {Pethick}, C.~J. 1981, Astrophys. J., 243, 1003

\bibitem[{{Finn} \& {Evans}(1990)}]{finn}
{Finn}, L.~S. \& {Evans}, C.~R. 1990, Astrophys. J., 351, 588

\bibitem[{{Flanagan} \& {Hughes}(1998)}]{flanagan}
{Flanagan}, {\'E}.~{\'E}. \& {Hughes}, S.~A. 1998, \prd, 57, 4566

\bibitem[{{Fryer}(2004)}]{fryersingle}
{Fryer}, C.~L. 2004, Astrophys. J. Lett., 601, L175

\bibitem[{{Fuller} {et~al.}(1985){Fuller}, {Fowler}, \& {Newman}}]{full85}
{Fuller}, G.~M., {Fowler}, W.~A., \& {Newman}, M.~J. 1985, Astrophys. J., 293,
  1

\bibitem[{{Harikae} {et~al.}(2010){Harikae}, {Kotake}, \&
  {Takiwaki}}]{harikae_b}
{Harikae}, S., {Kotake}, K., \& {Takiwaki}, T. 2010, Astrophys. J., 713, 304

\bibitem[{{Harikae} {et~al.}(2009){Harikae}, {Takiwaki}, \&
  {Kotake}}]{harikae_a}
{Harikae}, S., {Takiwaki}, T., \& {Kotake}, K. 2009, Astrophys. J., 704, 354

\bibitem[{{Heger} \& {Langer}(2000)}]{hege00}
{Heger}, A. \& {Langer}, N. 2000, Astrophys. J., 544, 1016

\bibitem[{{Hirata} {et~al.}(1987){Hirata}, {Kajita}, {Koshiba}, {Nakahata}, \&
  {Oyama}}]{hirata}
{Hirata}, K., {Kajita}, T., {Koshiba}, M., {Nakahata}, M., \& {Oyama}, Y. 1987,
  Physical Review Letters, 58, 1490

\bibitem[{{Hough} {et~al.}(2005){Hough}, {Rowan}, \& {Sathyaprakash}}]{hough}
{Hough}, J., {Rowan}, S., \& {Sathyaprakash}, B.~S. 2005, Journal of Physics B
  Atomic Molecular Physics, 38, 497

\bibitem[{{Itoh} {et~al.}(1989){Itoh}, {Adachi}, {Nakagawa}, {Kohyama}, \&
  {Munakata}}]{itoh89}
{Itoh}, N., {Adachi}, T., {Nakagawa}, M., {Kohyama}, Y., \& {Munakata}, H.
  1989, Astrophys. J., 339, 354

\bibitem[{{Itoh} {et~al.}(1990){Itoh}, {Adachi}, {Nakagawa}, {Kohyama}, \&
  {Munakata}}]{itoh90}
---. 1990, Astrophys. J., 360, 741

\bibitem[{{Janka} \& {Moenchmeyer}(1989)}]{jamu}
{Janka}, H. \& {Moenchmeyer}, R. 1989, \aap, 209, L5

\bibitem[{{Kawagoe} {et~al.}(2009){Kawagoe}, {Takiwaki}, \& {Kotake}}]{kawa}
{Kawagoe}, S., {Takiwaki}, T., \& {Kotake}, K. 2009, Journal of Cosmology and
  Astro-Particle Physics, 9, 33

\bibitem[{{Kawamura} {et~al.}(2006){Kawamura}, {Nakamura}, {Ando}, {Seto},
  {Tsubono}, {Numata}, {Takahashi}, {Nagano}, {Ishikawa}, {Musha}, {Ueda},
  {Sato}, {Hosokawa}, {Agatsuma}, {Akutsu}, {Aoyanagi}, {Arai}, {Araya},
  {Asada}, {Aso}, {Chiba}, {Ebisuzaki}, {Eriguchi}, {Fujimoto}, {Fukushima},
  {Futamase}, {Ganzu}, {Harada}, {Hashimoto}, {Hayama}, {Hikida}, {Himemoto},
  {Hirabayashi}, {Hiramatsu}, {Ichiki}, {Ikegami}, {Inoue}, {Ioka},
  {Ishidoshiro}, {Itoh}, {Kamagasako}, {Kanda}, {Kawashima}, {Kirihara},
  {Kiuchi}, {Kobayashi}, {Kohri}, {Kojima}, {Kokeyama}, {Kozai}, {Kudoh},
  {Kunimori}, {Kuroda}, {Maeda}, {Matsuhara}, {Mino}, {Miyakawa}, {Miyoki},
  {Mizusawa}, {Morisawa}, {Mukohyama}, {Naito}, {Nakagawa}, {Nakamura},
  {Nakano}, {Nakao}, {Nishizawa}, {Niwa}, {Nozawa}, {Ohashi}, {Ohishi},
  {Ohkawa}, {Okutomi}, {Oohara}, {Sago}, {Saijo}, {Sakagami}, {Sakata},
  {Sasaki}, {Sato}, {Shibata}, {Shinkai}, {Somiya}, {Sotani}, {Sugiyama},
  {Tagoshi}, {Takahashi}, {Takahashi}, {Takahashi}, {Takano}, {Tanaka},
  {Taniguchi}, {Taruya}, {Tashiro}, {Tokunari}, {Tsujikawa}, {Tsunesada},
  {Yamamoto}, {Yamazaki}, {Yokoyama}, {Yoo}, {Yoshida}, \&
  {Yoshino}}]{fpdecigo}
{Kawamura}, S., {Nakamura}, T., {Ando}, M., {Seto}, N., {Tsubono}, K.,
  {Numata}, K., {Takahashi}, R., {Nagano}, S., {Ishikawa}, T., {Musha}, M.,
  {Ueda}, K., {Sato}, T., {Hosokawa}, M., {Agatsuma}, K., {Akutsu}, T.,
  {Aoyanagi}, K., {Arai}, K., {Araya}, A., {Asada}, H., {Aso}, Y., {Chiba}, T.,
  {Ebisuzaki}, T., {Eriguchi}, Y., {Fujimoto}, M., {Fukushima}, M., {Futamase},
  T., {Ganzu}, K., {Harada}, T., {Hashimoto}, T., {Hayama}, K., {Hikida}, W.,
  {Himemoto}, Y., {Hirabayashi}, H., {Hiramatsu}, T., {Ichiki}, K., {Ikegami},
  T., {Inoue}, K.~T., {Ioka}, K., {Ishidoshiro}, K., {Itoh}, Y., {Kamagasako},
  S., {Kanda}, N., {Kawashima}, N., {Kirihara}, H., {Kiuchi}, K., {Kobayashi},
  S., {Kohri}, K., {Kojima}, Y., {Kokeyama}, K., {Kozai}, Y., {Kudoh}, H.,
  {Kunimori}, H., {Kuroda}, K., {Maeda}, K., {Matsuhara}, H., {Mino}, Y.,
  {Miyakawa}, O., {Miyoki}, S., {Mizusawa}, H., {Morisawa}, T., {Mukohyama},
  S., {Naito}, I., {Nakagawa}, N., {Nakamura}, K., {Nakano}, H., {Nakao}, K.,
  {Nishizawa}, A., {Niwa}, Y., {Nozawa}, C., {Ohashi}, M., {Ohishi}, N.,
  {Ohkawa}, M., {Okutomi}, A., {Oohara}, K., {Sago}, N., {Saijo}, M.,
  {Sakagami}, M., {Sakata}, S., {Sasaki}, M., {Sato}, S., {Shibata}, M.,
  {Shinkai}, H., {Somiya}, K., {Sotani}, H., {Sugiyama}, N., {Tagoshi}, H.,
  {Takahashi}, T., {Takahashi}, H., {Takahashi}, R., {Takano}, T., {Tanaka},
  T., {Taniguchi}, K., {Taruya}, A., {Tashiro}, H., {Tokunari}, M.,
  {Tsujikawa}, S., {Tsunesada}, Y., {Yamamoto}, K., {Yamazaki}, T., {Yokoyama},
  J., {Yoo}, C., {Yoshida}, S., \& {Yoshino}, T. 2006, Classical and Quantum
  Gravity, 23, 125

\bibitem[{{Kotake} {et~al.}(2009{\natexlab{a}}){Kotake}, {Iwakami}, {Ohnishi},
  \& {Yamada}}]{kotake_ray}
{Kotake}, K., {Iwakami}, W., {Ohnishi}, N., \& {Yamada}, S. 2009{\natexlab{a}},
  Astrophys. J., 704, 951

\bibitem[{{Kotake} {et~al.}(2009{\natexlab{b}}){Kotake}, {Iwakami}, {Ohnishi},
  \& {Yamada}}]{kotake09}
---. 2009{\natexlab{b}}, Astrophys. J. Lett., 697, L133

\bibitem[{{Kotake} {et~al.}(2011){Kotake}, {Iwakami-Nakano}, \&
  {Ohnishi}}]{kotake11}
{Kotake}, K., {Iwakami-Nakano}, W., \& {Ohnishi}, N. 2011, \apj, 736, 124

\bibitem[{{Kotake} {et~al.}(2007){Kotake}, {Ohnishi}, \& {Yamada}}]{kotake07}
{Kotake}, K., {Ohnishi}, N., \& {Yamada}, S. 2007, Astrophys. J., 655, 406

\bibitem[{{Kotake} {et~al.}(2006){Kotake}, {Sato}, \& {Takahashi}}]{kota06}
{Kotake}, K., {Sato}, K., \& {Takahashi}, K. 2006, Reports of Progress in
  Physics, 69, 971

\bibitem[{{Kotake} {et~al.}(2004{\natexlab{a}}){Kotake}, {Sawai}, {Yamada}, \&
  {Sato}}]{kota04b}
{Kotake}, K., {Sawai}, H., {Yamada}, S., \& {Sato}, K. 2004{\natexlab{a}},
  Astrophys. J., 608, 391

\bibitem[{{Kotake} {et~al.}(2003{\natexlab{a}}){Kotake}, {Yamada}, \&
  {Sato}}]{kota03a}
{Kotake}, K., {Yamada}, S., \& {Sato}, K. 2003{\natexlab{a}}, Astrophys. J.,
  595, 304

\bibitem[{{Kotake} {et~al.}(2003{\natexlab{b}}){Kotake}, {Yamada}, \&
  {Sato}}]{kotakegw}
---. 2003{\natexlab{b}}, \prd, 68, 044023

\bibitem[{{Kotake} {et~al.}(2005){Kotake}, {Yamada}, \& {Sato}}]{kotake05}
---. 2005, \apj, 618, 474

\bibitem[{{Kotake} {et~al.}(2004{\natexlab{b}}){Kotake}, {Yamada}, {Sato},
  {Sumiyoshi}, {Ono}, \& {Suzuki}}]{kota04a}
{Kotake}, K., {Yamada}, S., {Sato}, K., {Sumiyoshi}, K., {Ono}, H., \&
  {Suzuki}, H. 2004{\natexlab{b}}, \prd, 69, 124004

\bibitem[{{Kudoh} {et~al.}(2006){Kudoh}, {Taruya}, {Hiramatsu}, \&
  {Himemoto}}]{kudoh}
{Kudoh}, H., {Taruya}, A., {Hiramatsu}, T., \& {Himemoto}, Y. 2006, \prd, 73,
  064006

\bibitem[{{Kuroda} \& {the LCGT Collaboration}(2006)}]{lcgt}
{Kuroda}, K. \& {the LCGT Collaboration}. 2006, Classical and Quantum Gravity,
  23, 215

\bibitem[{{Leonor} {et~al.}(2010){Leonor}, {Cadonati}, {Coccia}, {D'Antonio},
  {Di Credico}, {Fafone}, {Frey}, {Fulgione}, {Katsavounidis}, {Ott},
  {Pagliaroli}, {Scholberg}, {Thrane}, \& {Vissani}}]{leonor}
{Leonor}, I., {Cadonati}, L., {Coccia}, E., {D'Antonio}, S., {Di Credico}, A.,
  {Fafone}, V., {Frey}, R., {Fulgione}, W., {Katsavounidis}, E., {Ott}, C.~D.,
  {Pagliaroli}, G., {Scholberg}, K., {Thrane}, E., \& {Vissani}, F. 2010,
  Classical and Quantum Gravity, 27, 084019

\bibitem[{{Liebend{\"o}rfer}(2005)}]{matthias05}
{Liebend{\"o}rfer}, M. 2005, \apj, 633, 1042

\bibitem[{{Marek} \& {Janka}(2009)}]{marek}
{Marek}, A. \& {Janka}, H.-T. 2009, Astrophys. J., 694, 664

\bibitem[{{Marek} {et~al.}(2009){Marek}, {Janka}, \& {M{\"u}ller}}]{marek_gw}
{Marek}, A., {Janka}, H.-T., \& {M{\"u}ller}, E. 2009, Astron. Astrophys., 496,
  475

\bibitem[{{Meszaros}(2006)}]{mesz06}
{Meszaros}, P. 2006, Reports of Progress in Physics, 69, 2259

\bibitem[{{M\"{o}nchmeyer} {et~al.}(1991){M\"{o}nchmeyer}, {Schaefer},
  {Mueller}, \& {Kates}}]{mm}
{M\"{o}nchmeyer}, R., {Schaefer}, G., {Mueller}, E., \& {Kates}, R.~E. 1991,
  Astron. Astrophys., 246, 417

\bibitem[{{M\"{u}ler} \& {Janka}(1997)}]{muyan97}
{M\"{u}ler}, E. \& {Janka}, H.-T. 1997, Astron. Astrophys., 317, 140

\bibitem[{{M{\"u}ller} {et~al.}(2004){M{\"u}ller}, {Rampp}, {Buras}, {Janka},
  \& {Shoemaker}}]{mueller04}
{M{\"u}ller}, E., {Rampp}, M., {Buras}, R., {Janka}, H.-T., \& {Shoemaker},
  D.~H. 2004, Astrophys. J., 603, 221

\bibitem[{{Murphy} {et~al.}(2009){Murphy}, {Ott}, \& {Burrows}}]{murphy}
{Murphy}, J.~W., {Ott}, C.~D., \& {Burrows}, A. 2009, \apj, 707, 1173

\bibitem[{{Obergaulinger} {et~al.}(2006{\natexlab{a}}){Obergaulinger}, {Aloy},
  {Dimmelmeier}, \& {M{\"u}ller}}]{ober06a}
{Obergaulinger}, M., {Aloy}, M.~A., {Dimmelmeier}, H., \& {M{\"u}ller}, E.
  2006{\natexlab{a}}, Astron. Astrophys., 457, 209

\bibitem[{{Obergaulinger} {et~al.}(2006{\natexlab{b}}){Obergaulinger}, {Aloy},
  \& {M{\"u}ller}}]{ober06b}
{Obergaulinger}, M., {Aloy}, M.~A., \& {M{\"u}ller}, E. 2006{\natexlab{b}},
  Astron. Astrophys., 450, 1107

\bibitem[{{Obergaulinger} {et~al.}(2009){Obergaulinger}, {Cerd{\'a}-Dur{\'a}n},
  {M{\"u}ller}, \& {Aloy}}]{MRI}
{Obergaulinger}, M., {Cerd{\'a}-Dur{\'a}n}, P., {M{\"u}ller}, E., \& {Aloy},
  M.~A. 2009, \aap, 498, 241

\bibitem[{{Ott}(2009)}]{ott_rev}
{Ott}, C.~D. 2009, Classical and Quantum Gravity, 26, 063001

\bibitem[{{Ott} {et~al.}(2006){Ott}, {Burrows}, {Dessart}, \&
  {Livne}}]{ott_new}
{Ott}, C.~D., {Burrows}, A., {Dessart}, L., \& {Livne}, E. 2006, Physical
  Review Letters, 96, 201102

\bibitem[{{Ott} {et~al.}(2008){Ott}, {Burrows}, {Dessart}, \&
  {Livne}}]{ott_angle}
---. 2008, \apj, 685, 1069

\bibitem[{{Ott} {et~al.}(2004){Ott}, {Burrows}, {Livne}, \& {Walder}}]{ott}
{Ott}, C.~D., {Burrows}, A., {Livne}, E., \& {Walder}, R. 2004, Astrophys. J.,
  600, 834

\bibitem[{{Ott} {et~al.}(2007{\natexlab{a}}){Ott}, {Dimmelmeier}, {Marek},
  {Janka}, {Hawke}, {Zink}, \& {Schnetter}}]{ott_prl}
{Ott}, C.~D., {Dimmelmeier}, H., {Marek}, A., {Janka}, H., {Hawke}, I., {Zink},
  B., \& {Schnetter}, E. 2007{\natexlab{a}}, Physical Review Letters, 98,
  261101

\bibitem[{{Ott} {et~al.}(2007{\natexlab{b}}){Ott}, {Dimmelmeier}, {Marek},
  {Janka}, {Zink}, {Hawke}, \& {Schnetter}}]{ott_2007}
{Ott}, C.~D., {Dimmelmeier}, H., {Marek}, A., {Janka}, H., {Zink}, B., {Hawke},
  I., \& {Schnetter}, E. 2007{\natexlab{b}}, Classical and Quantum Gravity, 24,
  139

\bibitem[{{Rosswog} \& {Liebend{\"o}rfer}(2003)}]{ross03}
{Rosswog}, S. \& {Liebend{\"o}rfer}, M. 2003, \mnras, 342, 673

\bibitem[{{Scheck} {et~al.}(2008){Scheck}, {Janka}, {Foglizzo}, \&
  {Kifonidis}}]{fog}
{Scheck}, L., {Janka}, H., {Foglizzo}, T., \& {Kifonidis}, K. 2008, \aap, 477,
  931

\bibitem[{{Scheck} {et~al.}(2004){Scheck}, {Plewa}, {Janka}, {Kifonidis}, \&
  {M{\"u}ller}}]{sche04}
{Scheck}, L., {Plewa}, T., {Janka}, H.-T., {Kifonidis}, K., \& {M{\"u}ller}, E.
  2004, Physical Review Letters, 92, 011103

\bibitem[{{Scheidegger} {et~al.}(2008){Scheidegger}, {Fischer}, {Whitehouse},
  \& {Liebend{\"o}rfer}}]{simon}
{Scheidegger}, S., {Fischer}, T., {Whitehouse}, S.~C., \& {Liebend{\"o}rfer},
  M. 2008, Astron. Astrophys., 490, 231

\bibitem[{{Scheidegger} {et~al.}(2010){Scheidegger}, {K{\"a}ppeli},
  {Whitehouse}, {Fischer}, \& {Liebend{\"o}rfer}}]{scheid}
{Scheidegger}, S., {K{\"a}ppeli}, R., {Whitehouse}, S.~C., {Fischer}, T., \&
  {Liebend{\"o}rfer}, M. 2010, \aap, 514, A51+

\bibitem[{{Shen} {et~al.}(1998){Shen}, {Toki}, {Oyamatsu}, \&
  {Sumiyoshi}}]{shen98}
{Shen}, H., {Toki}, H., {Oyamatsu}, K., \& {Sumiyoshi}, K. 1998, Nuclear
  Physics A, 637, 435

\bibitem[{{Shibata} {et~al.}(2006){Shibata}, {Liu}, {Shapiro}, \&
  {Stephens}}]{shib06}
{Shibata}, M., {Liu}, Y.~T., {Shapiro}, S.~L., \& {Stephens}, B.~C. 2006, \prd,
  74, 104026

\bibitem[{{Shibata} \& {Sekiguchi}(2004)}]{shibaseki}
{Shibata}, M. \& {Sekiguchi}, Y.-I. 2004, \prd, 69, 084024

\bibitem[{{Suwa} {et~al.}(2009){Suwa}, {Kotake}, {Takiwaki}, {Whitehouse},
  {Liebendoerfer}, \& {Sato}}]{suwa}
{Suwa}, Y., {Kotake}, K., {Takiwaki}, T., {Whitehouse}, S.~C., {Liebendoerfer},
  M., \& {Sato}, K. 2009, ArXiv e-prints

\bibitem[{{Takahashi} {et~al.}(1978){Takahashi}, {El Eid}, \&
  {Hillebrandt}}]{taka78}
{Takahashi}, K., {El Eid}, M.~F., \& {Hillebrandt}, W. 1978, Astron.
  Astrophys., 67, 185

\bibitem[{{Takiwaki} {et~al.}(2004){Takiwaki}, {Kotake}, {Nagataki}, \&
  {Sato}}]{taki04}
{Takiwaki}, T., {Kotake}, K., {Nagataki}, S., \& {Sato}, K. 2004, Astrophys.
  J., 616, 1086

\bibitem[{{Takiwaki} {et~al.}(2009){Takiwaki}, {Kotake}, \& {Sato}}]{taki09}
{Takiwaki}, T., {Kotake}, K., \& {Sato}, K. 2009, Astrophys. J., 691, 1360

\bibitem[{{Thorne}(1980)}]{thorne}
{Thorne}, K.~S. 1980, Reviews of Modern Physics, 52, 299

\bibitem[{{van Elewyck} {et~al.}(2009){van Elewyck}, {Ando}, {Aso}, {Baret},
  {Barsuglia}, {Bartos}, {Chassande-Mottin}, {di Palma}, {Dwyer}, {Finley},
  {Kei}, {Kouchner}, {Marka}, {Marka}, {Rollins}, {Ott}, {Pradier}, \&
  {Searle}}]{van}
{van Elewyck}, V., {Ando}, S., {Aso}, Y., {Baret}, B., {Barsuglia}, M.,
  {Bartos}, I., {Chassande-Mottin}, E., {di Palma}, I., {Dwyer}, J., {Finley},
  C., {Kei}, K., {Kouchner}, A., {Marka}, S., {Marka}, Z., {Rollins}, J.,
  {Ott}, C.~D., {Pradier}, T., \& {Searle}, A. 2009, International Journal of
  Modern Physics D, 18, 1655

\bibitem[{{Walder} {et~al.}(2005){Walder}, {Burrows}, {Ott}, {Livne},
  {Lichtenstadt}, \& {Jarrah}}]{walder}
{Walder}, R., {Burrows}, A., {Ott}, C.~D., {Livne}, E., {Lichtenstadt}, I., \&
  {Jarrah}, M. 2005, \apj, 626, 317

\bibitem[{{Weinstein}(2002)}]{advancedligo}
{Weinstein}, A. 2002, Classical and Quantum Gravity, 19, 1575

\bibitem[{{Woosley} \& {Bloom}(2006)}]{woos_blom}
{Woosley}, S.~E. \& {Bloom}, J.~S. 2006, \araa, 44, 507

\bibitem[{{Woosley} \& {Heger}(2006)}]{woos06}
{Woosley}, S.~E. \& {Heger}, A. 2006, Astrophys. J., 637, 914

\bibitem[{{Yamada} \& {Sawai}(2004)}]{yama04}
{Yamada}, S. \& {Sawai}, H. 2004, Astrophys. J., 608, 907

\bibitem[{{Yoon} \& {Langer}(2005)}]{yoon}
{Yoon}, S.-C. \& {Langer}, N. 2005, Astron. Astrophys., 443, 643

\bibitem[{{Zwerger} \& {Mueller}(1997)}]{zweg}
{Zwerger}, T. \& {Mueller}, E. 1997, Astron. Astrophys., 320, 209

\end{thebibliography}

\end{document}